\documentclass{article}
\usepackage[utf8]{inputenc}
\usepackage{graphicx}
\usepackage{color}
\usepackage{subcaption}
\usepackage{amssymb}
\usepackage{mathtools}
\usepackage{amsmath} 
\usepackage{dsfont}
\usepackage[numbers]{natbib} 
\usepackage{multirow}
\usepackage{caption}
\usepackage{subcaption}

\usepackage{MnSymbol}
\usepackage{lscape}
\usepackage{rotating}
\usepackage{array}
\usepackage{siunitx}

\usepackage{float}
\usepackage{placeins}
\usepackage{dutchcal}
\usepackage{lineno}
\usepackage{hyperref}

\usepackage[]{algorithm2e}
\usepackage{framed}
\definecolor{shadecolor}{gray}{0.95}

\definecolor{blueberry}{RGB}{0,0,153}
\definecolor{TODO}{RGB}{100,10,150}
\usepackage[justification=centering]{caption}
\definecolor{PhD}{RGB}{3,78,60}
\definecolor{copied}{RGB}{35,139,60}
 \definecolor{light-gray}{gray}{0.9}
 \definecolor{draft}{RGB}{4, 90, 141}
  \definecolor{done}{RGB}{35,139,69}
  \definecolor{maybe}{RGB}{250,0,50}

\makeatletter
\renewcommand\@fnsymbol[1]{%
  \ifcase#1\or 1\or 2\or 3\or 4 \fi%
}
\makeatother

\newcommand{\Xicum}{X_i^{\text {cum }}}

\newcommand{\wpt}{w(p_t)}
\newcommand{\indep}{\perp\!\!\!\perp}

\title{Addressing complex structures of measurement error arising in the exposure assessment in occupational epidemiology using a Bayesian hierarchical approach}
\author{
  Raphael Rehms\thanks{Institute for Medical Information Processing, Biometry and Epidemiology, Faculty of Medicine, LMU Munich, Germany}\textsuperscript{~,*}  \and
  Nicole Ellenbach\footnotemark[1]\textsuperscript{~,}\thanks{Munich Center for Machine Learning (MCML)} \and
  Veronika Deffner\thanks{Federal Office for Radiation Protection, Germany} \and
  Sabine Hoffmann\thanks{Department of Statistics, LMU Munich, Germany}
}

\begin{document}

\maketitle

\begingroup
\makeatletter
\renewcommand\@fnsymbol[1]{%
  \ifcase#1\or *\or \dagger\or \ddagger\or
  \mathsection\or \mathparagraph\or \|\or **\or \dagger\dagger\else\@ctrerr\fi%
}
\renewcommand{\thefootnote}{\fnsymbol{footnote}}
\footnotetext[1]{Corresponding author: rrehms@ibe.med.uni-muenchen.de}
\makeatother
\endgroup

\section*{Abstract}
Exposure assessment in occupational epidemiology may involve multiple unknown quantities that are measured or reconstructed simultaneously for groups of workers and over several years. Additionally, exposures may be collected using different assessment strategies, depending on the period of exposure. As a consequence, researchers who are analyzing occupational cohort studies are commonly faced with challenging structures of exposure measurement error, involving complex dependence structures and multiple measurement error models, depending on the period of exposure. However, previous work has often made many simplifying assumptions concerning these errors. In this work, we propose a Bayesian hierarchical approach to account for a broad range of error structures arising in occupational epidemiology. The considered error structures may involve several unknown quantities that can be subject to mixtures of Berkson and classical measurement error. It is possible to account for different error structures, depending on the exposure period and the location of a worker. Moreover, errors can present complex dependence structures over time and between workers. We illustrate the proposed hierarchical approach on a subgroup of the German cohort of uranium miners to account for potential exposure uncertainties in the association between radon exposure and lung cancer mortality. The performance of the proposed approach and its sensitivity to model misspecification are evaluated in a simulation study. The results show that biases in estimates arising from very complex measurement errors can be corrected through the proposed Bayesian hierarchical approach.


\section{Introduction}
\label{sec:inroduction}

  In occupational epidemiology, researchers are often interested in the association between the cumulative exposure to a specific chemical or physical agent and the time until an event occurs, such as a diagnosis of or death from a certain disease.
 In this situation, exposure is time-dependent and ongoing, 
  and the exposure history of workers may be collected using different assessment strategies depending on the period of exposure. During the estimation process, measurement errors can arise, leading to complex patterns of exposure uncertainty, where the structure and magnitude of measurement error can vary over time. \\
In many occupational cohort studies, there are no prospective exposure measurements.
As it is often infeasible or too costly to measure or to estimate exposure values for individual workers, many occupational cohort studies rely on job-exposure matrices (JEMs). In general, JEMs provide information about exposure levels for certain job categories or titles \citep{Greenland2015}.  The assigned exposure value may also vary with respect to location and year.
 Although it is widely acknowledged that exposure measurement error can have deleterious consequences on the validity of statistical inference and may lead to erroneous conclusions, researchers in the field of occupational epidemiology do not often account for these errors \citep{burstyn2020occupational}.  Many of the publications account for measurement error use Bayesian approaches \citep{belloni2020bayesian, singer2018bayesian, burstyn2018correction, bartell2017bayesian, Hoffmann2017,little2014association}, while frequentist methods are also employed \citep{Greenland2015, little2020analysis, little2014association, allodji2012performance}.  It is often assumed that errors follow a simple structure where deviations of individual exposures for workers from the assigned exposure level in a JEM is described as unshared Berkson error \citep{armstrong1998,bender2005, kuchenhoff2007, Steenland2015}. However, the estimation of the exposure values in a JEM often involves the estimation of multiple uncertain quantities that may be subject to complex structures of measurement error. 
  Exposure values in a JEM are often reconstructed retrospectively by experts \citep{Peters2020}, leading to measurement errors that may affect several groups of workers and several exposure years at the same time. 
 As a consequence, group-level estimates and individual job conditions  may give rise to mixtures of Berkson and classical measurement error with  complex dependency structures. While it is difficult to account for complex error structures using available measurement error correction methods, they pose serious threats to the validity of statistical inference in occupational epidemiology. Moreover, it is unlikely that the bias introduced by these complex measurement error structures can be adequately corrected for with methods that assume simple measurement error structures, such as unshared Berkson error.  In previous work, we found
 that uncertainty components shared within workers cause more bias in risk estimation than components of unshared exposure uncertainty and that  this can lead to an attenuation of the exposure-response curve for high exposure values \citep{Hoffmann2018}, 
   a phenomenon that is frequently observed in occupational cohort studies \citep{HertzPicciotto1993,Stayner2003,Steenland2015}.  \\
The aim of this paper is to demonstrate that even highly complex measurement error structures, which may commonly arise in occupational cohort studies, can be effectively addressed. We illustrate this by proposing a flexible hierarchical Bayesian framework capable of accounting for these complexities, applied to the German cohort of uranium miners.
Section 2 describes measurement error structures that typically arise through the prospective and retrospective exposure assessment in occupational epidemiology and the basic methodology to account for them using a Bayesian hierarchical approach. In section 3, we demonstrate the flexibility of the approach as a proof of concept using the German cohort of uranium miners as illustrative example. We account for highly complex structures of measurement errors that arises when studying the association between radon exposure and lung cancer mortality, presenting preliminary results for a selective subgroup of the Wismut cohort \citep{Kreuzer2010, Kuchenhoff2018}. 
Section 4 presents a simulation study in which we evaluate the performance of the proposed approach and assess its sensitivity to model misspecification. In section 5 we discuss our results and give an outlook for future work.

\section{Accounting for measurement error in occupational epidemiology} \label{sec:measurement_error_correction_general} 
\subsection{Measurement error characteristics in occupational epidemiology}
When describing measurement error, one commonly distinguishes classical and Berkson error. A classical measurement error model describes the error prone observed value $Z$ as a function of the true value $X$ and of an error term $U$ that is independent of $X$. For additive and multiplicative error, classical measurement error can be written as  $Z = X + U$ and  $Z = X \cdot U$  with $X \indep U$, respectively. Conversely, for Berkson error, the error term U is independent of the observed value $Z$. Again, we can write $X = Z + U$ and  $X = Z \cdot U$  with $Z \indep U$ for additive and multiplicative Berkson error, respectively. 
A classical measurement error model is often employed to describe the measurement arising from a measurement device or through the estimation of experts whereas a Berkson model describes the situation where the true and unknown value of a quantity of interest deviates from a fixed and observed value. 
In an occupational cohort study, it is in general plausible to assume that exposure measurement error is non-differential, i.e. that errors are independent of the outcome since it is unlikely that errors arising in the exposure estimation in an occupational cohort depend on the (future) disease status of individual workers.\\
 If a JEM is used,  the same exposure level is typically assigned to all workers in a job category (potentially also as a function of year and location). 
As a consequence, measurement errors that arise from the estimation of this common exposure level will affect all workers in that job category in the same way. In cases where exposure values in a JEM are based on measurements and there is only one quantity that is to be measured (for instance pesticide levels, airborne contaminants or noise) the error arising through the estimation of the exposure level given in the JEM can be described through a classical measurement error component that is shared among workers in the same job category: $Z(t, j) = X(t, j) + U(t, j)$. Here, every worker who works at year $t$ in job category $j$ will receive the same error $ U(t, j)$.
 In cases where exposure values are reconstructed by experts rather than being based on measurements, the classical error component described above will often be shared for the entire time period for which the estimation was made, leading to a classical measurement error that is shared among workers and years: $Z(p_t, j) = X(p_t, j) + U(p_t, j)$ for all job categories $j$ and years $t$ in the time period $p_t$.  \\
 Deviations of each individual true exposure from the common exposure level can then be described by unshared Berkson error. 
If the same value is assumed for different locations, the true average value at each location may deviate from this common value, leading to a Berkson error that is shared among all workers in a given location. For instance, an additive shared Berkson error at different locations $o$ (e.g. mining objects) and years $t$ could be written as  $X(t, o) = Z(t, o)  + U (t, o)$.
Finally, even when exposure values are estimated in a prospective fashion, there may be additional uncertain quantities involved in the estimation that are retrospectively estimated by experts. We will describe more complex error structures involving several uncertain quantities in more detail in our application to the German cohort of uranium miners.

\subsection{Accounting for measurement error}
To account for measurement error using a Bayesian hierarchical approach, one specifies three sub-models and concatenates them to a full joint model using conditional independence assumptions \citep{richardson1993bayesian}. This is similar to the procedure in a likelihood-based approach \citep{carroll2012measurement, Gustafson2004}. However, Bayesian approaches may be more versatile from a computational perspective when it comes to the correction for complex errors structures. 
In the following, we will give a short summary of the general Bayesian approach. We start by defining the three required sub-models: The disease model, measurement model and the exposure model.\\[1mm]
\textbf{Disease model:} The disease model defines the relation $[Y|X,\theta_1]$ between an outcome $Y$ and one or more covariates of interest $X$ (at least exposure values), where $\theta_1$ is the collection of all parameters of the disease model. We follow Richardson and Gilks \cite{richardson1993bayesian}, to denote (possibly conditional) distributions using squared brackets. In occupational cohort studies, a commonly used outcome is the time until an event occurs, for instance the time until a specific diagnosis or cause-specific death where the covariate of interest would be an exposure to a specific chemical or physical agent. \\[1mm]
\textbf{Measurement model:} 
The measurement model $[Z|X, \theta_2]$ describes the relation between the observed, error prone variable $Z$ and the true, unobserved values $X$ for all uncertain quantities intervening in the calculation of individual exposure values, parameterized by $\theta_2$. With respect to measurement errors that arise through application of a JEM, a combination of a classical and Berkson error can be used to describe the following situation: In general, the estimation of the exposure for one group or job category is not measured precisely. Therefore, a classical error can be assumed to reflect uncertainty in the estimation of this common exposure level. Moreover, an additional Berkson error can describe deviations of exposure values of individual workers from this common exposure level. This leads to a combination of two different errors. Since multiple uncertain quantities typically intervene in the estimation of an exposure level in a JEM, we may often be faced with combinations of Berkson and classical measurement error in multiple uncertain quantities. We illustrate this situation in more detail in the next section for the Wismut cohort.\\[1mm]
\textbf{Exposure model:} The exposure model defines the distribution $[X|\theta_3]$ of the unobserved exposure $X$. Note, that in the case of a Berkson error, a formulation of an exposure model is not required. $\theta_3$ parameterized the distribution of $X$. If we assume, for instance, that exposure values follow a normal distribution, the parameters would be $\theta_3=(\mu_X, \sigma_X^2)$, i.e the assumed mean value and standard deviation of the distribution of $X$. \\

We can combined the three models using conditional independence assumptions to formulate the unnormalized joint posterior over all unknown quantities by assuming additional prior distributions on the parameters of the models $[\theta_1]$, $[\theta_2]$ and $[\theta_3]$:
\begin{equation} \label{eq:posterior_generic}
    [\theta_1, \theta_2, \theta_3, X| Y, Z] \propto [\theta_1][\theta_2][\theta_3][Y|X, \theta_1][Z|X, \theta_2][X|\theta_3].
\end{equation}

We can use any suitable inference method to obtain the posterior. Most prominent choices are Markov chain Monte Carlo (MCMC) \citep{brooks2011}, variational inference \citep{blei2017variational} or integrated nested Laplace approximation \citep{Muff2015,rue2009approximate}. MCMC can be considered as the most versatile approach and it can approximate the posterior with arbitrary accuracy (at least in theory). This usually comes with a substantially higher computational burden. MCMC algorithms generate a Markov chain that has the posterior of interest as stationary distribution. After initializing the chain at an arbitrary state, the chain will converge to the stationary distribution (samples before convergence are typically discarded as burnin). Samples from this Markov chain can then be considered as samples from the posterior. It is common to have a calibration phase at the beginning to sample more efficiently and to thin the chain to obtain less correlated samples. Furthermore, it is good practice to run multiple chains to parallelize computation and to calculate common quality criteria like the $\hat{R}$ statistic \citep{gelman1992inference, vehtari2021rank}.

\section{Application: Modeling the association between radon exposure and lung cancer mortality in the Wismut cohort}
\label{sec:application}

In the following, we will use the German cohort of uranium miners, also referred to as Wismut cohort \citep{Kreuzer2010}, as an example to illustrate how complex structures of potential exposure uncertainties that may arise in occupational cohort studies can be accounted for through a Bayesian hierarchical approach. The cohort consists of 58,974 workers who were employed between 1946 and 1989 at the Wismut company. 
We are interested in the association between the exposure of radon gas (or rather the decay products) and lung cancer mortality. It is generally acknowledged that the exposure to radon progeny is a relevant cause of lung cancer \citep{UNSCEAR2020}.
For all exposure years in the cohort, individual exposure estimates were based on a JEM \citep{Lehmann1998, Lehmann2004}, which provides estimated annual exposure values to radon progeny for a reference activity with 2000 working hours. These values were then multiplied by a so-called activity weighting factor that can be summarized as a correction factor for the different radiation exposures associated with different activities. Further, the estimated annual exposure is multiplied by  a working time factor to adjust for deviations in the number of standard  working hours from a reference \citep{Kuchenhoff2018}. The individual annual exposure is afterwards calculated by combining these quantities (see section \ref{sec: ME_modelM2} that explains this in more detail as part of the measurement model). \\

\subsection{Disease model}
In the Wismut cohort, we are interested in the association between radon exposure and lung cancer mortality. We use a survival outcome, where lung cancer mortality is a right-censored variable $(Y_i, \delta_i)$ for each miner $i \in \{1, \dots n\}$, where $Y_i$ denotes the attained age in years and $\delta_i$ is the censoring indicator. Attained age is left truncated (at the time of entry into the cohort) and radon exposure, denoted as $\Xicum(t)$, is a time-varying covariate that accumulates over years $t$. Two popular model choices in radiation epidemiology are the proportional hazards (PH) and the excess hazard ratio (EHR) model. The instantaneous hazard for miner $i$ in a PH model is defined as 

\begin{equation*}
    h_i(t ; \boldsymbol\lambda, \beta)= h_0(t, \boldsymbol\lambda)\exp (\beta \cdot \Xicum(t) ),
\end{equation*}
where $h_0$ is the baseline hazard and $\beta$ the effect of the exposure on the instantaneous hazard. The baseline hazard does not depend on the covariates, but may depend on time and a set of parameters $\boldsymbol{\lambda}$. In the case of the EHR model, th e instantaneous hazard $h_i(t; \boldsymbol\lambda, \beta)$ is modeled as 

\begin{equation*}
    h_i(t ; \boldsymbol\lambda, \beta)= h_0(t, \boldsymbol\lambda) (1 + \beta \cdot \Xicum(t)).
\end{equation*}

The EHR model implies a constraint on $\beta$ as the hazard must be positive. 
In both cases, we assume a simplified linear model without effect-modifying variables and  we model the baseline hazard assuming an \textit{explicitly modeled functional form}
as the correction for measurement error through a hierarchical model requires the formulation of the full likelihood. 
We choose a flexible piecewise-constant function as model for the hazard baseline,  parameterized through $\boldsymbol\lambda = \{\lambda_1, \dots, \lambda_4\}$, i.e.  $h_0(t, \boldsymbol{\lambda})=\lambda_k \quad \forall t \in I_k=\left(s_{k-1}, s_k\right]$, where $I_k$ is the time interval corresponding to the baseline hazard of $\lambda_j$ that is defined through the partitions $0=s_0 < s_1<s_2<s_3<s_4$ \citep{ibrahim2001bayesian, Fahrmeir2003, Martino2011, Majumdar2007}. Following Hoffmann et al. \cite{Hoffmann2017}, we use $s_1=40, s_2=55, s_3=75, s_4=104$ as break points and define the priors on the parameters as
\begin{equation*}
    \beta \sim N(0, \sigma_\beta^2)
\end{equation*}
\centerline{and}
\begin{equation*}
    \lambda_k \sim Ga(\alpha^\lambda_k, \beta^\lambda_k) \qquad \forall k \in \{1,\dots 4\},
\end{equation*}
where each $\lambda_k$ has an individual prior specification through shape and scale parameters $\alpha^\lambda_k$ and $\beta^\lambda_k$ to reflect a stepwise increase in the baseline hazard. The prior on $\beta$ is chosen to be uninformative while the parameters of the baseline hazard are chosen to be informative (see first part of section \ref{sec:appendix:priors} in the supplementary for a list of all chosen prior parameters of the disease model).

\subsection{Measurement model for the Wismut cohort} \label{sec: ME_modelM2}
\begin{figure}[ht]
    \centering
    \includegraphics[width=0.9\linewidth]{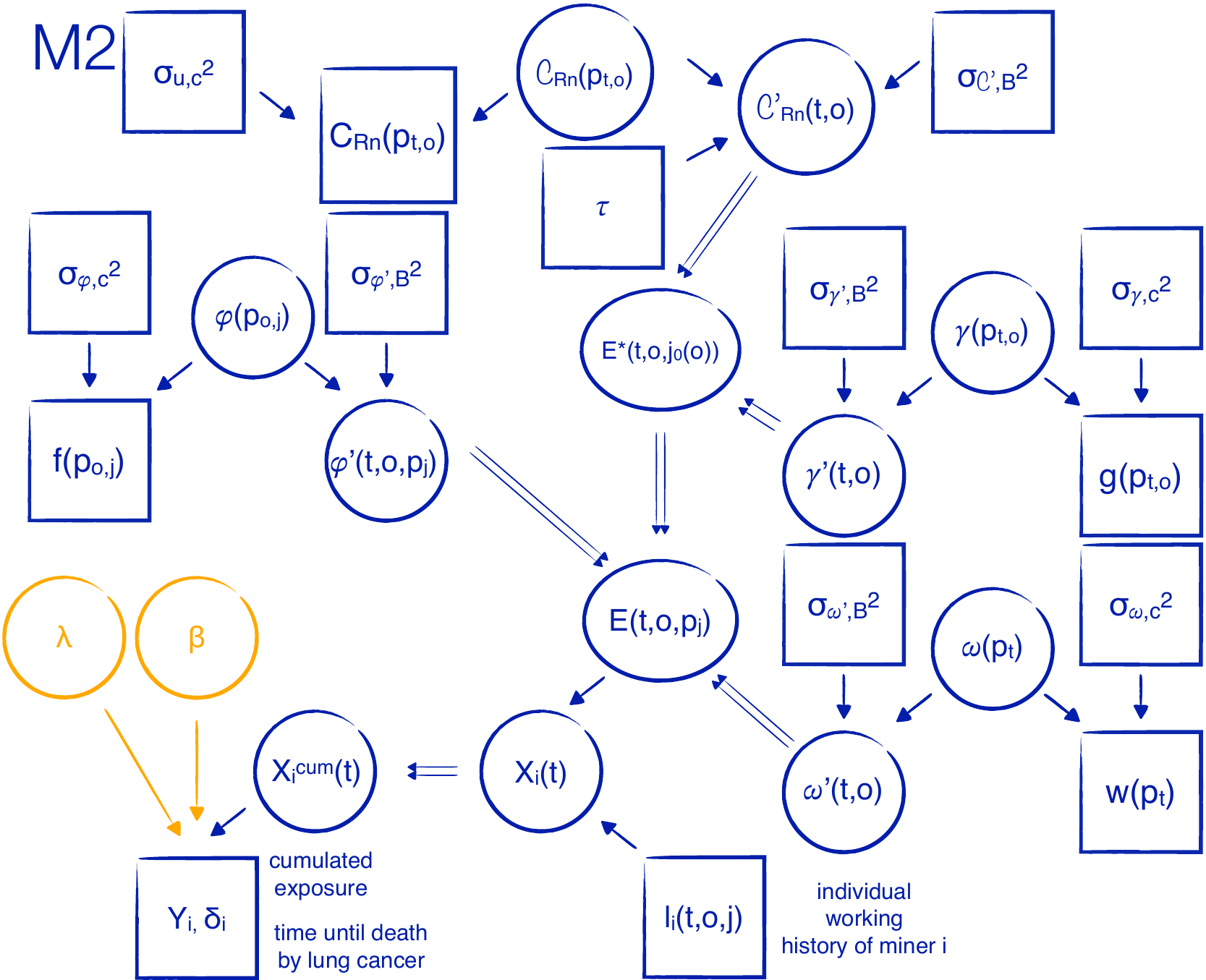}
    \caption{DAG of measurement model M2. With circles we represent unknown quantities and with boxes observed/fixed quantities. Single arrows indicate probabilistic dependencies between the different quantities and double arrows deterministic dependencies. The DAG does not include the prior parameters and parameters of the exposure model.}
    \label{fig:DAG_M2}
\end{figure}

In the Wismut cohort, the exposure values in the JEM were estimated through different methods 
depending on the time period and the type of workplace (underground, open pit, milling or surface). A mining location is also referred to as an object in the Wismut cohort. Based on the preliminary work of Küchenhoff et al. \cite{Kuchenhoff2018}, Ellenbach et al. \cite{ellenbach2023ermittlung} characterize, quantify and develop measurement models to describe the characteristics of exposure uncertainties arising in the exposure assessment for all time periods and types of workplaces (see Figure \ref{fig:Wismut_objects} in section \ref{sec:appendix:ME_models_all} for an overview). We consider five of these different measurement models for the Bayesian approach: M1, M2, M2\_Expert and M3 for underground miners (depending on the time period and the availability of measurements for radon gas and radon progeny), and M4 for miners employed at surface areas affiliated to mining locations. Besides, we assume no measurement error for miners working in pure surface objects without exposure to radon. We do not consider the error structures arising in processing companies and in open pit mining objects and exclude miners who ever worked in either of these two types of mining locations, leading to a sub sample of 48,534 miners. Furthermore, we exclude all miners whose working histories include measurement model MX\_Expert\_WLM, which was defined by Ellenbach et al.\cite{ellenbach2023ermittlung} for cases in which information to reconstruct the exposure values according to the other measurement models was lacking, finally leading to a selective sub sample of 34,809 miners. The cumulated exposure of a miner is often derived considering different measurement models, since most miners worked in more than just one time period or changed the location over time. 
For the sake of readability, in this section, we describe the measurement model M2 as it represents a typical error structure that may arise through the use of a JEM. We refer to the supplementary material \ref{sec:appendix:ME_models_all} for the full specification of all measurement models. \\ 
Measurement model M2 was employed for workers in underground mining  objects located in the federal states Saxony and Thuringia, Germany and development objects in Saxony in the exposure assessment period 1955/56 to 1965 in Saxony and  1955/56-1974 in Thuringia. In this exposure assessment period, exposure values were calculated using the following formula: 
\begin{equation} \label{eq:M2true}
    E(t, o, j)=12 \cdot C_{R n}(p_{t,o}) \cdot \tau(t,o) \cdot f(p_{o, j}) \cdot w\left(p_t\right) \cdot g\left(p_{t, o}\right),
\end{equation}
where $E(t, o, j)$ denotes the estimated annual exposure to radon for a worker who conducted activity $j$ in location $o$  and year $t$. $12 \cdot C_{R n}(p_{t,o}) \cdot \tau(t,o)$ is the estimated annual radon gas concentration for the reference activity (being a hewer) and 2000 annual working hours.
To obtain the annual exposure to radon progeny, the exposure is multiplied by an activity weighting factor $f(p_{o, j})$, which corrects for the fact that most activities had lower exposure to radon than a hewer, as well as by a working time factor $\wpt$, which modifies the reference working time of 2000 hours to obtain a smaller or higher amount of working hours. By multiplying with an equilibrium factor $g\left(p_{t, o}\right)$, the measured radon gas exposure is converted to radon progeny exposure in working level months (WLM), which is the historical unit of radon exposure in cohorts of uranium miners and related to the potential alpha energy concentration \citep{Marsh2012}. We call these different quantities \textit{uncertain factors} as they are considered to be potentially error-prone.
With a slight abuse of notation, we define variables $p_t$, $p_{t,o}$ and $p_{o,j}$ to express the dependence structures arising from the fact that many of the uncertain factors were not estimated for individual years $t$ and locations $o$, but instead a common value was used for several years, locations and activities (e.g. $p_{t,o}$ uses a common value for several years $t$ and locations $o$). 
Although there were ambient radon gas concentration measurements $C_{R n}(p_{t,o})$ available for most years and locations in measurement model M2, 
there are some years and locations for which there were no measurements. For these cases, measurements from different years or locations were extrapolated and sometimes adjusted with a transfer factor $\tau(t,o)$.
In general, the use of a common value or extrapolated measurements for several years, locations, and activities leads to a shared classical measurement error for these years, locations, and activities.\\
For the radon gas concentrations, we assume a classical error component that describes potential uncertainty in the measurement process. Since the average radon gas concentration is the result of a large number of measurements, we assume that the average of the measurements is distributed normally around its true value, following the central limit theorem. 
 Additionally, we assume a Berkson error component only for those years and location without ambient radon gas measurements but used the transferred values from other years or locations. 
 For the activity weighting, the working time and the equilibrium factors, we assume both a classical and a Berkson measurement error component to describe the potential uncertainty in the estimation of a common value and the variability around this common value for several years, objects and activities, respectively:
\begin{align*}
& C_{R n}(p_{t, o}) =\mathcal{C}_{R n}(p_{t, o})+U_{\mathcal{C}, c}(p_{t, o})       \\
& \mathcal{C}_{R n}^{\prime}(t, o) =\mathcal{C}_{R n}(p_{t, o}) \cdot U_{\mathcal{C}_{R n}^{\prime}, B}(t,o)   \cdot \tau (t,o)     & \text{(only if values were transferred) } \\[6pt]
& f(p_{o, j}) =\varphi(p_{o, j}) \cdot U_{\varphi, c}(p_{o, j})          \\
& \varphi^{\prime}(t, o, j) =\varphi(p_{o, j}) \cdot U_{\varphi^{\prime}, B}(t, o, j)          \\[6pt]
& w\left(p_t\right) =\omega\left(p_t\right) \cdot U_{\omega, c}\left(p_t\right)           \\
& \omega^{\prime}(t, o) =\omega\left(p_t\right) \cdot U_{\omega{ }^{\prime}, B}(t, o)            \\[6pt]
& g\left(p_{t, o}\right) =\gamma\left(p_{t, o}\right) \cdot U_{\gamma, c}\left(p_{t, o}\right)           \\
& \gamma^{\prime}(t, o) =\gamma\left(p_{t,o}\right) \cdot U_{\gamma^{\prime}, B}(t, o)            \\
\end{align*}

where $\mathcal{C}_{R n}^{\prime}(t, o)$, $\varphi^\prime(t, o,j)$,  $\omega^{\prime}(t, o)$, and $\gamma^{\prime}(t, o)$ are the true values of the radon gas concentration, the activity weighting factor, the  working time factor, and the equilibrium factor respectively.
$C_{R n}(p_{t, o})$, $\varphi(p_{o, j})$, $\omega\left(p_t\right)$ and $\gamma\left(p_{t, o}\right)$ are the true average (level) values for each of them and
$C_{R n}(p_{t, o})$, $f(p_{o, j})$, $w\left(p_t\right)$ and $g\left(p_{t, o}\right)$ are the level values that were estimated by experts.
For additive errors, we assume an error term that follows a normal distribution, i.e. $U_{\mathcal{C}, c}(p_{t, o}) \sim N(0, \sigma^2_{\mathcal{C},c}(p_{t, o}))$, for the radon gas concentration while we assume a log-normal distributed error for multiplicative errors. That is 
$\log(U_{\mathcal{C}_{R n}^{\prime}, B}(t, o)) \sim N\left(-\frac{1}{2}\sigma^2_{\mathcal{C}_{R n}^{\prime}, B}(t, o), \sigma^2_{\mathcal{C}_{R n}^{\prime}, B}(t, o)\right)$ for the Berkson error of the radon gas concentration.
We assume that the error distributions for the classical and the Berkson error components of the other uncertain factors also follow a log-normal distribution analogous to the Berkson error of the radon gas concentration.\\
The \textit{true} exposure of a miner $i$ who is employed in activity $j$ in object $o$ and year $t$ is then given by: 

\begin{equation} \label{eq:M2Berkson}
    X_i(t, o, j)=12 \cdot \mathcal{C}^\prime_{R n}(t,o)  \cdot \varphi^{\prime}(t, o, p_j) \cdot \omega^{\prime}(t, o) \cdot \gamma^{\prime}(t, o) \cdot l_i(t,o,j) 
\end{equation}

\noindent where the factor $l_i(t,o,j)$ accounts for the individual time that a  miner worked in object $o$ and activity $j$ in year $t$. 
Equation (\ref{eq:M2Berkson}) resembles equation (\ref{eq:M2true}). However, it is important to note that now the different factors are \textit{latent variables} that are estimated in addition to the parameters of interest.\\
Given the individual exposure $X_i$, a cumulated individual exposure $X^\text{cum}_i$ must be calculated as exposure accumulates over time. For details and an example, see section \ref{sec:update_latent_expo} of the supplementary material. We show the full measurement error model M2 in Figure \ref{fig:DAG_M2} as a directed acyclic graph (DAG). See Ellenbach et al. \cite{ellenbach2023ermittlung} for the DAGs of the other measurement models.

\subsection{Exposure model for the Wismut cohort}
In the Wismut cohort, the uncertain factors vary depending on the measurement model. We describe the exposure models for measurement model M2 and refer to section \ref{sec:appendix:priors} of the supplementary material for the exposure models of the other measurement models. For M2, we do not only need to define exposure models for the true and unknown radon gas concentration $\mathcal{C}_{R n}(p_{t, o})$, but also for the other uncertain factors, namely the activity weighting factor $\varphi^\prime(t, o,j)$, the  working time factor  $\omega^{\prime}(t, o)$ and the equilibrium factor $\gamma^{\prime}(t, o)$. 
The radon gas concentration is assumed to follow a truncated Gaussian distribution with varying priors over time on the mean and standard deviation to reflect varying concentration levels over the years, i.e.

\begin{align*}
    \mathcal{C}_{R n}(p_{t, o}) & \sim N^+(\mu_{\mathcal{C}_{Rn}}(t), \sigma^2_{\mathcal{C}_{Rn}}(t)) \\
    \mu_{\mathcal{C}_{Rn}}(t) & \sim N(\mu_{\mathcal{C}_{Rn}}, \sigma_{\mathcal{C}_{Rn}}^2) \\
    \sigma_{\mathcal{C}_{Rn}}(t) & \sim N^+(\mu_{\mathcal{C}_{Rn}}, \sigma_{\mathcal{C}_{Rn}}^2).
\end{align*}

The other uncertain factors $\varphi(p_{o, j}), \omega\left(p_t\right)$, and $\gamma\left(p_{t, o}\right)$ are assumed to follow a generalized Beta distribution with modified support to an appropriate range that represents realistic values:

\begin{align*}
    \varphi(p_{o, j}) & \sim B_{[0,1.3]}(a_\varphi, b_\varphi), \\
    \omega\left(p_t\right) & \sim B_{[0.6,1.5]}(a_\omega, b_\omega), \\
    \gamma\left(p_{t, o}\right) & \sim B_{[0.05,0.8]}(a_\gamma, b_\gamma),
\end{align*}

where we use $B_{[lo,up]}(a, b)$ to denote a Beta distribution with parameters $a$ and $b$ and write the modified support in subscript with $[lo, up]$ and we allow for additional flexibility by  
setting the following priors on the shape parameters of the Beta distributions ($a$ and $b$):

\begin{align*}
    a_m, b_m \sim N(\mu_{B_m}, \sigma_{B_m}^2), \quad \text{for} \quad m \in \{ \varphi, \omega , \gamma\}.
\end{align*}

\subsection{Inference using an efficient MCMC algorithm}
\label{sec:inference}
After defining the disease, measurement and exposure model, we can use them to write the unnormalized joint posterior for all workers and years where the exposure assessment is based on measurement model M2. To get a compact form, we use again squared brackets to denote the probability density function (PDF) of a random variable.

\begin{align*}
& {[\boldsymbol{\theta}, X^{cum} \mid \cdot]} \propto \\[0.25cm]
& {[\beta][\boldsymbol\lambda]\left[a_\omega\right]\left[b_\omega\right]\left[a_\gamma\right]\left[b_\gamma\right]\left[a_{\varphi}\right]\left[b_{\varphi}\right] \prod_t [\mu_{\mathcal{C}_{Rn}}(t)] \prod_t [\sigma_{\mathcal{C}_{Rn}}(t)] \times } \\
& \prod_{i, t}\left[Y_i \mid \boldsymbol\lambda, \beta, \Xicum(t)\right] \times \\
& \prod_{i, t}\left[\Xicum(t) \mid X_i(t) \right] \times \\
& \prod_{i, t}\left[X_i(t) \mid \mathcal{C}^\prime_{R_n}(t, o), \varphi^{\prime}(t, o, j), \gamma^{\prime}(t, o), \omega^{\prime}(t, o), l_i(t,o,j) \right] \times \\
& \prod_{t, o}\left[\omega^{\prime}(t, o) \mid \sigma_{\omega^\prime, B}^2, \omega\left(p_t\right)\right] \prod_{p_t}\left[w\left(p_t\right) \mid \sigma_{\omega, c}^2, \omega\left(p_t\right)\right] \prod_{p_t}\left[\omega\left(p_t\right) \mid a_\omega, a_\omega\right] \times \\
& \prod_{t, o}\left[\gamma^{\prime}(t, o) \mid \sigma_{\gamma^{\prime}, B}^2, \gamma\left(p_{t, o}\right)\right] \prod_{p_{t, o}}\left[g\left(p_{t, o}\right) \mid \sigma_{\gamma, c}^2, \gamma\left(p_{t, o}\right)\right] \prod_{p_{t, o}}\left[\gamma\left(p_{t, o}\right) \mid a_\gamma, a_\gamma\right] \times \\
& \prod_{t,o,j}\left[\varphi^{\prime}(t, o, j) \mid \sigma_{\varphi^{\prime}, B}^2, \varphi(p_{o, j})\right] \prod_{p_{o, j}}\left[f(p_{o, j}) \mid \sigma_{\varphi, c}^2, \varphi(p_{o, j})\right] \prod_{p_{o, j}}\left[\varphi(p_{o, j}) \mid a_{\varphi}, a_{\varphi}\right] \times \\
& \prod_{t, o}\left[\mathcal{C}^\prime_{R_n}(t, o) \mid \sigma_{\mathcal{C}, c}^2, \mathcal{C}_{R_n}(p_{t, o})\right] \prod_{p_{t, o}}\left[\mathcal{C}_{R_n}(p_{t, o}) \mid \mu_{\mathcal{C}_{Rn}}(t), \sigma^2_{\mathcal{C}_{Rn}}(t)\right]
\end{align*}
where $\boldsymbol{\theta}$ denotes the collection of all latent quantities, i.e. $\boldsymbol{\theta}$ = $(\beta$, $\boldsymbol\lambda$, $a_\omega$, $b_\omega$, $a_\gamma$, $b_\gamma$, $a_\varphi$, $b_\varphi$, $\mu_{\mathcal{C}}(t)$, $\sigma_{\mathcal{C}}(t)$, $\mathcal{C}^\prime_{R_n}(t, o)$, $\varphi^{\prime}(t, o, j)$, $\gamma^{\prime}(t, o)$, $\omega^{\prime}(t, o)$, $\mathcal{C}_{R_n}(p_{t, o})$, $\varphi(p_{o, j})$, $\gamma\left(p_{t, o}\right)$, $\omega\left(p_t\right))$. 
Note that the fourth and fifth line do not represent a probabilistic, but a \textit{deterministic} relationship that expresses the connection between the individual cumulated exposure of a worker and the value of the uncertain quantities intervening in the calculation of the yearly exposure values.\\ 
We implement a custom Metropolis-Hastings (MH) algorithm with component-wise updates \citep{Metropolis1987, brooks2011} to sample from the posterior. We briefly describe the key-points of the MCMC algorithm. More details are presented in section \ref{sec:appendix:update_scheme_long} in the supplementary material. 
All uncertain quantities are treated as latent variables that are updated at each sampling step of the MCMC algorithm one at a time, starting with the parameters of the disease model. For the parameters of the disease model, it is possible to condition on the cumulative latent exposure $\Xicum$ leading to a simplified MH-ratio where only the disease model has to be evaluated. 
The update of the latent exposure poses more challenges: After proposing a new value for one of the uncertain factors, formula (\ref{eq:M2Berkson}) can be used to calculate the latent exposure and to evaluate the disease model. However, this requires two extra steps: \\
1) Because the quantities are affected by shared classical and Berkson measurement error within and between workers, caution is necessary as each error is shared across different domains defined through the dependency structures specified in the measurement model. For instance, the classical error part of the working time factor $\omega$, varies only \textit{over periods $p_t$} (i.e. it is shared for multiple years), while the Berkson error is defined over \textit{every year $t$ and object $o$}. As a consequence, the uncertain factor has to be $mapped$ to its corresponding domain before it can be used for the calculation of the true and unknown exposure values.\\
2) Given a new proposed state, the calculation in (\ref{eq:M2Berkson}) returns only annual exposure values. However, as the exposure of a worker accumulates over time, it is necessary to calculate the cumulative exposure vector for each individual worker over all the working years every time a yearly exposure value is proposed. \\
We used sparse matrix multiplication to solve both challenges. This is computationally efficient because only non-zero values are stored and used for the mapping and cumulation.\\
We implement the MCMC update scheme in \texttt{python3} \citep{python3} in an object-oriented fashion using mainly the standard numerical library \texttt{numpy} \citep{numpy}. Statistical distributions and sparse matrix functionalities rely on \texttt{scipy} \citep{scipy}.

\subsection{Results}
\begin{figure}[ht]
    \centering
        \makebox[\textwidth][c]{\includegraphics[width=1.2\textwidth]{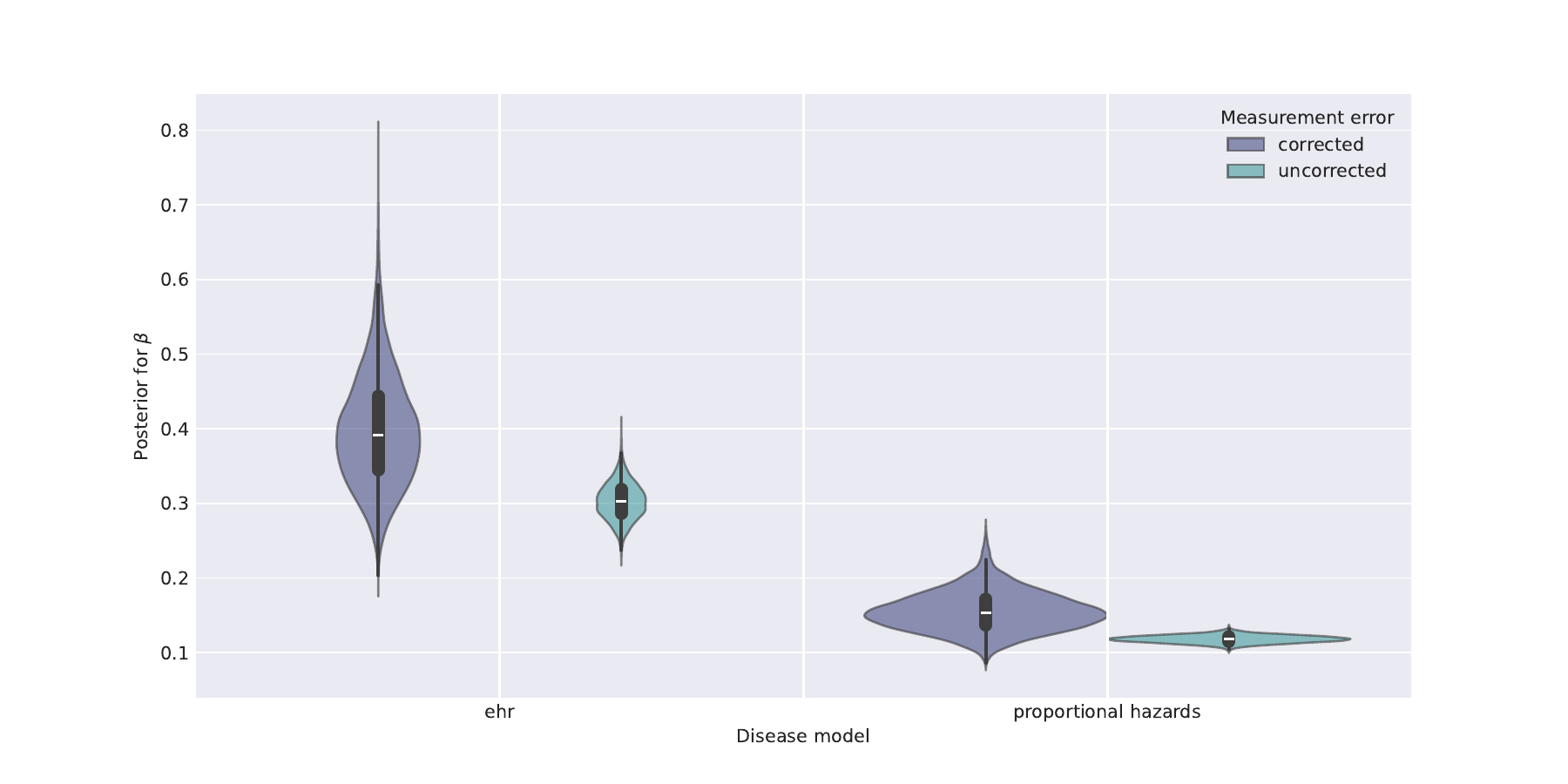}}
    \caption{Violin plots of the samples from the posterior distribution for an EHR (left) and proportional hazards model (right). Results that account for measurement error are plotted in blue and the results from the naive models without measurement error correction are plotted in green.}
    \label{fig:res_application}
\end{figure}

In order to obtain 4000 independent samples from the posterior distribution, we generate samples from eight independent chains with 100,000 iterations each and and thin them by keeping only every 200th sample).   
Beforehand, we tune the chains using 100 adaptive phases with 50 samples each to obtain better sampling quality and run further 50,000 iterations as burnin. Figure \ref{fig:res_application} and Table \ref{tab:results_application} show the results for a proportional hazards and an EHR model. We present as point estimates the empirical mean and median and the 95\%-highest density interval (credible interval) as measure of uncertainty. The HDI represents the 95\% of the most credible values \citep{kruschke2014doing}.

\begin{table}[ht]
\centering
\begin{tabular}{|c|c|c|c|c|}
\hline
\textbf{Disease model} & \textbf{Correction} & \textbf{Mean} & \textbf{Median} & \textbf{HDI (95\%)} \\
\hline
\multirow{2}{*}{Proportional hazards} & Corrected & 0.1553 & 0.1532 & [0.1066, 0.2032] \\
 & Uncorrected & 0.1181 & 0.1181 & [0.1088, 0.1275] \\
\hline
\multirow{2}{*}{EHR} & Corrected & 0.3977 & 0.3917 & [0.2614, 0.5352] \\
 & Uncorrected & 0.3028 & 0.3025 & [0.2567, 0.3471] \\
\hline
\end{tabular}
\caption{Summary statistics of the parameter $\beta$ (association between radon exposure and lung cancer mortality) using a proportional hazards or an excess hazards (EHR) model for the application on data of a selective subgroup of the German uranium miners cohort.  We present the empirical mean, median and HDI calculated from the sampled posterior.}
\label{tab:results_application}
\end{table}

It is observable that without accounting for the assumed measurement error structure, the estimated association between radon exposure and lung cancer mortality in the selective subgroup of the cohort is underestimated by about 23.95\% for the proportional hazards model and 23.86\% for the EHR model (when considering the empirical mean of the posterior distribution as point estimate). This indicates that in this example, accounting for measurement error results in an increase in the point estimate of the risk. However, it also considerably increases the uncertainty of the risk estimate. 
We provide the convergence analysis in section \ref{sec:appendix:convergence_application} of the supplementary material.

\section{Simulation study}
We conduct a simulation study to assess whether the proposed Bayesian hierarchical approach can produce reliable results when accounting for complex structures of measurement error that may typically arise in occupational cohorts. In the following, we will follow the structure proposed in Morris et al. \cite{morris2019using}.

\subsubsection*{Aims}
1) We want to ensure that an unbiased estimate for the parameter of interest can be inferred; 2) We aim to test the frequentist properties of the proposed approach by verifying if the 95\% credible interval from the posterior samples adequately covers the parameter of interest under the correct model assumptions; 3) We want to assess the sensitivity to model misspecification by estimating the extent to which an incorrect specification of the distributions of the measurement model can influence the results. \\

\subsubsection*{Data-generating mechanisms}
We generate 100 data sets per scenario, where each data set simulates the working history and survival times of 5,000 miners. The simulation study uses \textit{not} only measurement model M2, but  generates data over \textit{all}  measurement models that were used for the application to the real data of the Wismut cohort, only with the small simplification of no reference objects and thus no Berkson error component for radon measurements. To generate data sets that follow the assumed probabilistic models of the different measurement models (see section \ref{sec:appendix:ME_models_all} of the supplementary materials), we first randomly draw 5000 miners from a simplified cohort data set (due to data protection reasons) using only the information on whether a miner worked at the Wismut company and whether he was exposed to radon in the respective year. All miners are randomly sampled into different objects and different activities. Secondly, we sample the true average values of all uncertain factors, as well as all their classical and Berkson errors from the respective distributions. We take the dependency structures into account and generate shared errors accordingly. For example, for the working time factor, we sample as many true mean values $\omega\left(p_t\right)$ from a Beta distribution and as many multiplicative classical errors $U_{\omega, c}\left(p_t\right)$ from a log-normal distribution, as there are different values for $p_t$. For the multiplicative Berkson errors $U_{\omega{ }^{\prime}, B}(t, o) $, we sample a separate value for each year $t$ and each object $o$ from a log-normal distribution. We then obtain the true values of the uncertain factors by multiplying the sampled true average values with the sampled Berkson errors (e.g., $\omega^{\prime}(t, o) =\omega\left(p_t\right) \cdot U_{\omega{ }^{\prime}, B}(t, o)$). By multiplying (or adding in the case of an additive measurement error) the sampled classical errors with the sampled true average values, we obtain the observed values of the uncertain factors (e.g., $w\left(p_t\right) =\omega\left(p_t\right) \cdot U_{\omega, c}\left(p_t\right) $).  Then, the miners' true exposures and their error-prone observed exposures are calculated using, respectively, the true or observed values of the uncertain factors according to the formula for the respective measurement model (see section \ref{sec:appendix:ME_models_all} of the supplementary materials). The Bayesian hierarchical approach uses the observed exposures for measurement error correction and for the uncorrected naive estimate. The true exposures, on the other hand, are used to generate the survival times. In particular, we generate the censored time until death by lung cancer according to a PH model as a function of a miner's radon exposure in WLM as time-varying covariate using a method that relies on the generation of truncated piecewise exponential random variables, initially proposed in Zhou \cite{Zhou2001} and further extended by Hendry \cite{Hendry2014} and Montez-Rath et al.
\cite{Montez2017}. For the exact implementation of the simulation code in \texttt{R} \citep{Rlang}, we refer to the accompanying git repository.

\subsubsection*{Estimand}
Our estimand is the parameter of interest $\beta$ representing the association between radon exposure and lung cancer mortality. In particular, we consider the samples drawn for $\beta$ as estimates of the posterior distribution. 

\subsubsection*{Methods}
We define three different scenarios for the simulation study. The first scenario (S1) simulates data assuming $\beta = 0.3$, whereas the second scenario (S2) uses $\beta=0.6$. For both scenarios we apply the proposed model with measurement error (ME) correction, as well as a naive one without ME correction to the respective simulated data. Both scenarios should test the correctness of the approach covering aims 1) and 2). The third scenario (S3) is designed for aim 3): We test the robustness of the model against wrong distributional assumptions by investigating the impact of assuming a log-normal distribution for the radon concentration measurements for those measurement models where the data is simulated using  a (truncated) normal distribution and vice versa assuming a (truncated) normal distribution for models where the data is simulated using a log-normal distribution. Furthermore, we want to investigate whether a misspecification of the distributional assumptions for the exposure models specified for the uncertain factors other than radon concentration (e.g. working time) impacts model performance in a significant way: Instead of flexible Beta distributions (with additional priors on $a$ and $b$), we force the model to use a fixed uniform distribution implying $a = b = 1$ for the latent factors while using the standard data generating process. All scenarios are fitted using solely a proportional hazards model for the exposure-disease relationship and no EHR to keep the computational cost feasible. For every scenario, we generated 100 data sets. For scenarios S1 and S2 we used the Bayesian approach exactly as described in section \ref{sec:application}, and for scenario S3 we only modified the exposure models to account for the wrong distributional assumptions. For scenario S1 and S2, we also run a model on the true, unknown values of the uncertain factors, as they were measured without any error. For this we solely use formula (\ref{eq:M2Berkson}) to calculate the exposure and use only the disease model. It can therefore be seen as a reference where one would expect very accurate estimates.
Due to convergence problems for some data set in the second scenario (S2), only 99 or 97 data sets are used for S2 (see section \ref{sec:appendix:convergence_application} in the supplementary materials). 

\subsubsection*{Performance measures} 
The main performance measure is the bias between the point estimate (empirical mean) of the posterior distribution and the true value of $\beta$. We quantify it by calculating the absolute and the relative bias. Furthermore, we estimate the mean squared error (MSE). We calculate these quantities using the empirical mean of the posterior distribution as point estimate $\hat{\beta}$. Our secondary performance measure is the proportion of the coverage of the true $\beta$ value (that was used to generate the data) in the interval estimate of $\beta$, estimated by the empirical 95\%-HDI.

\subsubsection*{Results} 
\begin{figure}[h!]
\vspace{-0.25cm}
     \centering
     \begin{subfigure}[b]{0.9\textwidth}
         \centering
         \includegraphics[width=\textwidth]{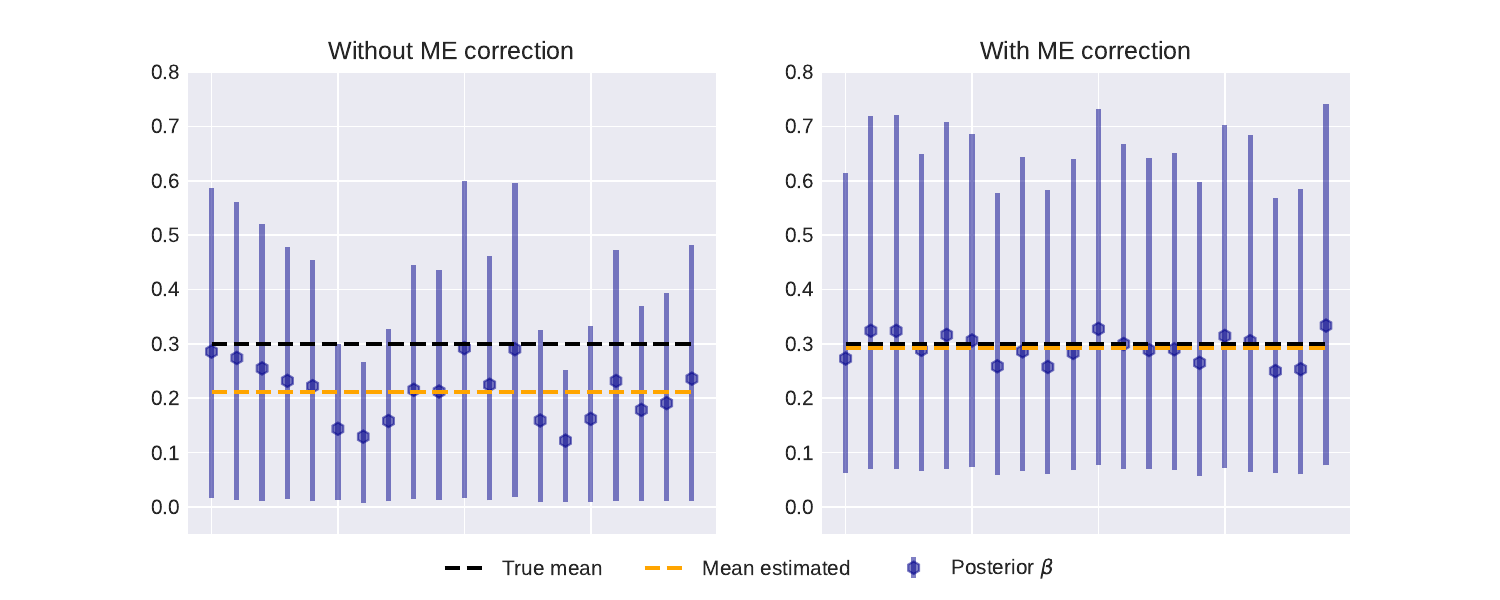}
         \caption{$\beta = 0.3$, left: naive, right: ME corrected}
     \end{subfigure}
         
     \vspace{-0.1cm}
     
     \begin{subfigure}[b]{0.9\textwidth}
         \centering
         \includegraphics[width=\textwidth]{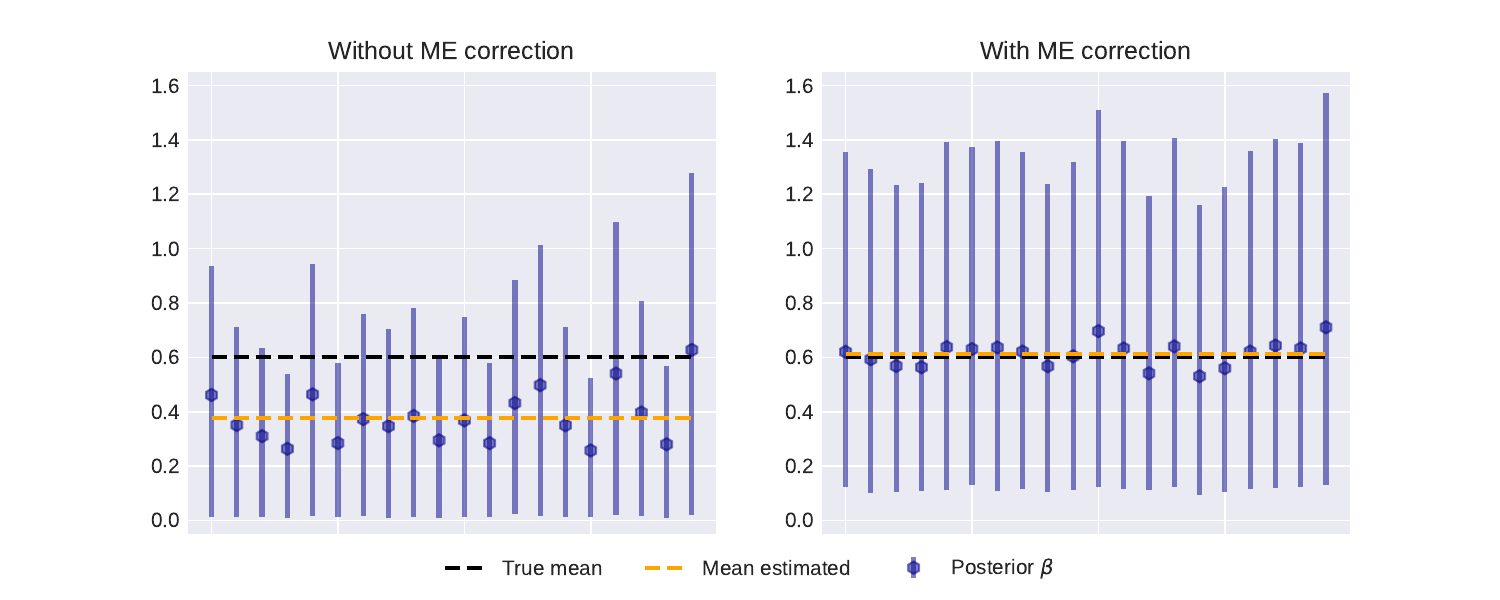}
         \caption{$\beta = 0.6$, left: naive, right: ME corrected}
     \end{subfigure}
     
     \vspace{-0.1cm}
     
     \begin{subfigure}[b]{0.9\textwidth}
         \centering
         \includegraphics[width=\textwidth]{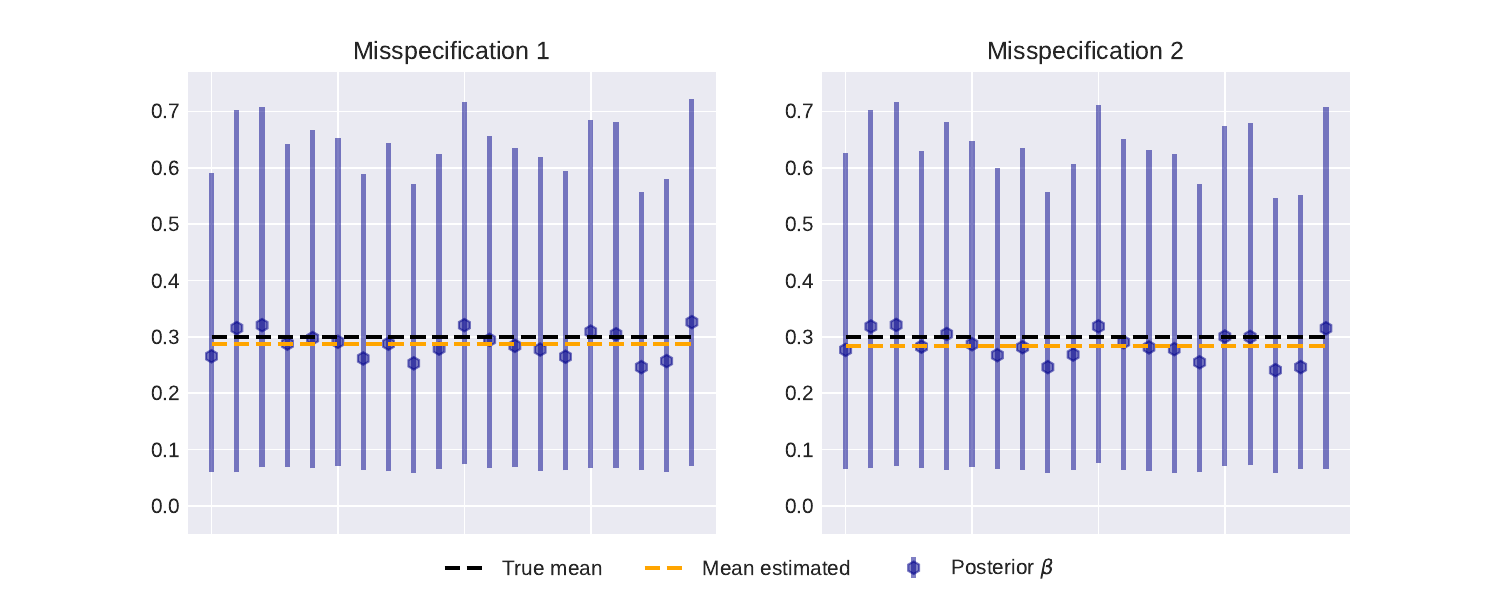}
         \caption{$\beta = 0.3$, left: misspecification first setting, right: misspecification second setting, both with ME correction}
     \end{subfigure}
        \caption{Empirical mean and 95\%-HDI (blue) derived from the posterior on the 20 first data sets of the simulation study. Horizontal line: mean value, yellow is the estimated empirical mean over all posterior means of $\beta$ and black denotes the true value of $\beta$.}
        \label{fig:results_simulation_study_20ds}
\end{figure}

Table \ref{tab:results_simulationstudy} shows the results for the different scenarios with their Monte Carlo standard error in parentheses, that quantifies the simulation uncertainty due to using a finite number of simulations \citep{morris2019using}. The results are in line with the results of the application on the data of the Wismut cohort: ignoring the measurement error may lead to some bias. Through measurement error correction, this bias can be eliminated for both scenarios S1 and S2 (aim 1) achieving a bias level that is nearly as good as fitting a model directly to the true values without measurement error. However, when accounting for measurement errors, the 95\%-HDI for the risk estimate becomes wider, even leading to overcoverage in scenarios with $beta=0.3$. Moreover, the results for S3 imply that a potential misspecification with respect to the exposure distribution on the radon gas measurements or on other uncertain factors has only a negligible impact on the estimates (aim 3). 
Further, we show the results of the posterior estimates for the first 20 simulated data sets graphically over all considered scenarios in Figure \ref{fig:results_simulation_study_20ds} (naive estimates and measurement error correction). Looking at the mean and 95\%-HDI, one can see that the measurement error correction provides good results while having slightly wider intervals caused by the higher uncertainty induced through the error (aim 3). In section \ref{sec:appendix:results_simulation_study} of the supplementary material, we analyze and discuss the convergence of the presented results.

\begin{table}[ht]
\centering
\resizebox{\textwidth}{!}{%
\begin{tabular}{l|ccc|ccc|cc|}
\cline{2-9}

                                                     & \multicolumn{3}{c|}{S 1 ($\beta = 0.3$)}                          & \multicolumn{3}{c|}{S 2 ($\beta = 0.6$)}                          & \multicolumn{2}{c|}{S 3 ($\beta = 0.3$)}                          \\
                                                     & \multicolumn{1}{l}{naive} & \multicolumn{1}{l}{ME cor.} & \multicolumn{1}{l|}{true} & \multicolumn{1}{l}{naive} & \multicolumn{1}{l}{ME cor.} & \multicolumn{1}{l|}{true} & \multicolumn{1}{l}{misspec1} & \multicolumn{1}{l|}{misspec2} \\ \hline
\multicolumn{1}{|c|}{\multirow{2}{*}{Absolute bias}} & -0.08                     & 0.004                        & -0.002                        & -0.207                    & 0.022                        & -0.003                        & -0.002                       & -0.004                        \\
\multicolumn{1}{|c|}{}                               & (0.005)                    & (0.003)                       & (0.001)                      & (0.01)                     & (0.006)                       & (0.001)                      & (0.003)                       & (0.003)                         \\ \hline
\multicolumn{1}{|l|}{\multirow{2}{*}{Relative bias}} & -0.266                    & 0.014                        & -0.007                        & -0.344                    & 0.037                        & -0.005                        & -0.007                       & -0.013                        \\
\multicolumn{1}{|l|}{}                               & (0.018)                    & (0.011)                       & (0.002)                      & (0.017)                    & (0.009)                       & (0.001)                      & (0.01)                        & (0.01)                          \\ \hline
\multicolumn{1}{|l|}{\multirow{2}{*}{MSE}}           & 0.0092                    & 0.001                        & 0.0001                        & 0.0529                    & 0.0036                       & 0.0001                        & 0.0009                       & 0.0009                        \\
\multicolumn{1}{|l|}{}                               & (0.001)                    & (0.0001)                      & (0.0)                      & (0.0043)                   & (0.0006)                      & (0.0)                      & (0.0001)                      & (0.0001)                          \\ \hline
\multicolumn{1}{|l|}{\multirow{2}{*}{Coverage}}      & 0.12                      & 0.99                         & 0.95                        & 0.031                     & 0.959                        & 0.959                        & 0.98                         & 1.0                           \\
\multicolumn{1}{|l|}{}                               & (0.325)                    & (0.099)                       & (0.218)                      & (0.172)                    & (0.198)                       & (0.199)                      & (0.14)                        & (0.0)                           \\ \hline
\end{tabular}%
}
\caption{Absolute/relative bias, mean squared error (MSE) and the coverage using 95\%-HDIs over 100 data sets (99 or 97 for S2, see section \ref{sec:appendix:results_simulation_study} in the supplementary materials) for the different scenarios, i.e. S1 with $\beta = 0.3$ , S2 with $\beta = 0.6$ and S3 with two misspecified models using also $\beta=0.3$. 'misspec1' assumes a log-normal distribution for the radon concentration measurements while data is simulated with a (truncated) normal or vice versa depending on the measurement model. 'misspec2' assumes a uniform distribution in the exposure model on other multiplicative factors. The column with 'true' calculates the model without measurement error correction on the true unobserved values and is therefore a reference model.}
\label{tab:results_simulationstudy}
\end{table}

\section{Discussion}
\label{sec:discussuin}
In this work, we proposed a Bayesian hierarchical approach to account for complex structures of measurement error in occupational cohort studies. These error structures can involve both multiple potentially uncertain quantities that may be subject to complex mixtures of Berkson and classical measurement error and multiple measurement error models to tailor the assumed measurement error structures to the exposure assessment strategies that were used for different workers and at different exposure periods. 
We illustrated the  approach on data of the Wismut cohort where and showed on simulated data that the proposed approach is able to produce reliable results under the assumed data generating processes.\\
However, a number of limitations have to be considered in the interpretation of our results. Like any statistical method, the proposed approach to account for measurement error may stand and fall with its implicitly and explicitly stated assumptions. In the simulation study, we investigated how assuming an additive error when the error is actually multiplicative and vice versa would affect our results. We tested the robustness to this misspecification of the measurement model, as a broad body of literature suggests a multiplicative error \citep{Lubin2005, Stram1999, Heid2002, Heid2004, Heidenreich2004uncertainties, heidenreich2012lung, allodji2012assessment, allodji2012impact, allodji2012performance} while we chose an additive error for measurements of radon gas concentration and radon progeny whenever multiple measurements were averaged.\\
We only considered a simplified model for the association between one (time-varying) exposure and an outcome, but ignored effect modifying variables that are known to be important in the association between radon exposure and lung cancer mortality. This requires more future work and was not the focus of this paper. Hence, the presented results should be interpreted as a proof of concept and illustration and rather not as an answer to the question what the actual effect of the variable of interest is. However, the used Bayesian hierarchical model would provide enough flexibility, to account for potential confounding and effect modifying variables in future work.

\subsection*{Conflicts of Interest}
The authors declare no conflicts of interest.

\subsection*{Data and code availability}
We provide the full code of the implemented MCMC sampler. We also provide all required files to re-run the simulation study.  Moreover, we provide the code that was used to run the application on the Wismut data. The repository can be found at \href{https://github.com/RaphaelRe/Wismut_ME_Bayes}{\url{https://github.com/RaphaelRe/Wismut_ME_Bayes}}. The actual data of the Wismut cohort cannot be shared due to privacy protection. However, we share the generated Markov chains for both, application and simulation \cite{rehms_2025_15050372} that can be used to produce all presented results.

\appendix

\newpage 
\section{Measurement models}
\label{sec:appendix:ME_models_all}

\begin{figure}[ht]
    \centering
    \makebox[\textwidth][c]{\includegraphics[width=\textwidth]{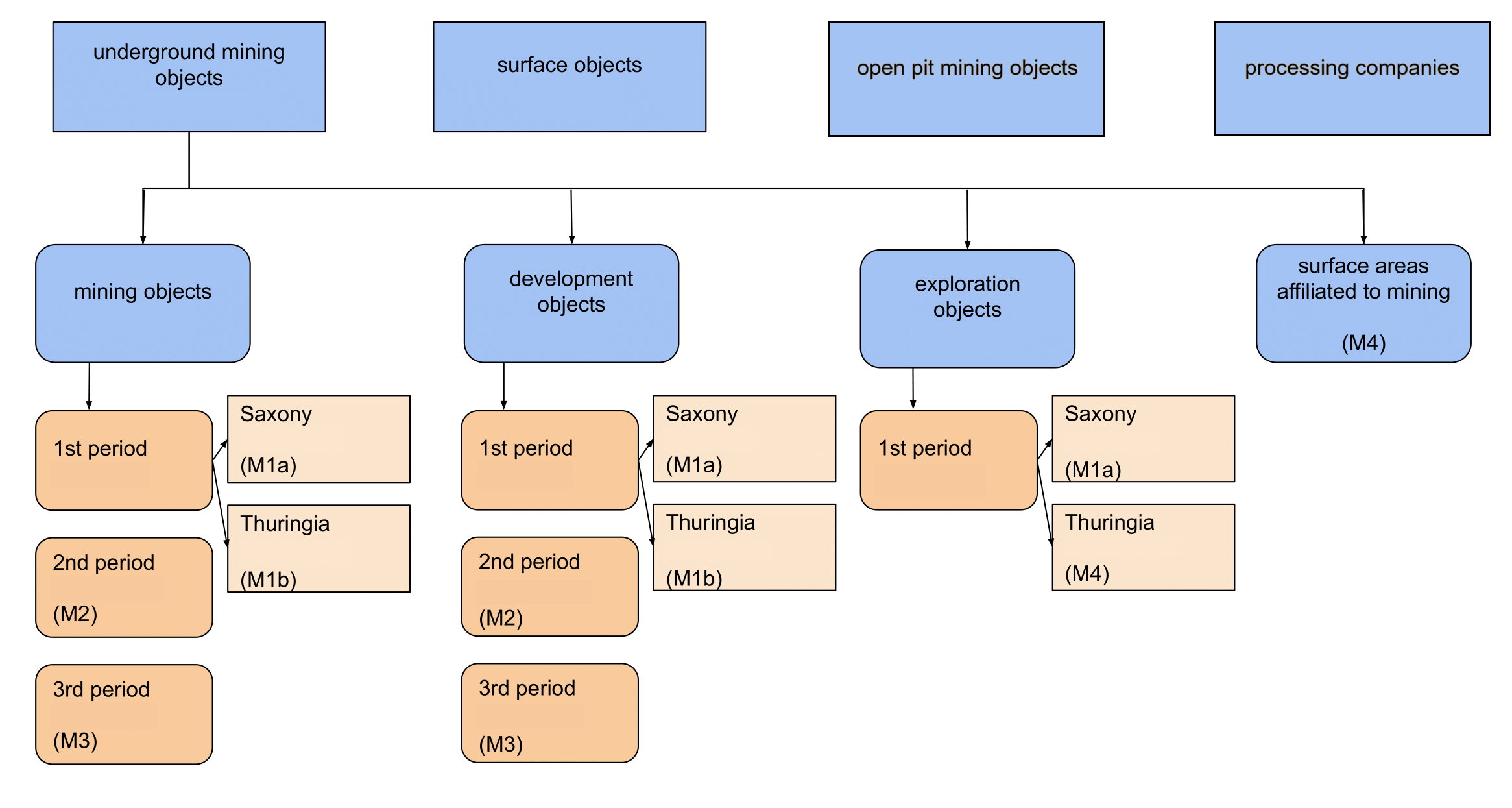}}
    \caption{The object structure of the Wismut cohort over the different time periods with the respective measurement models. Open pit mining objects and processing companies are not used for the Bayesian approach. Note that miners in surface objects are assumed to have no radon exposure and no measurement error and are thus shown without a measurement model.}
    \label{fig:Wismut_objects}
\end{figure}

In this section, we show the considered measurement models, that are used for the Bayesian hierarchical approach. Figure \ref{fig:Wismut_objects} is an adapted version from  Ellenbach et al.\cite{ellenbach2023ermittlung} and shows the object structure of the Wismut cohort and the respectively assumed measurement models. We now give the formulas for the considered measurement models. Multiplicative measurement errors always follow a log-Normal distribution while additive errors follow a Normal distribution. 

\subsubsection*{M1a}

Formula for calculation of exposure:

\begin{align*}
X_i(t,o,j)= & \left(\mathcal{C}_{Rn}(1937/1938,003) \cdot \mathcal{b}'(t,o) + \right.\\
            &  r(t,o) \cdot \frac{\mathcal{C}_{Rn}(t_0(o_0(o)), o_0(o))}{A(t_0(o_0(o)), o_0(o))} \cdot \tau_e'(t,o) \cdot A(t,o)) \cdot \\
            & \gamma'(t,o) \cdot \omega'(t,o) \cdot \varphi'(t,o,p_j) \cdot 12 \cdot  l_i(t,o,j) \cdot \tau(t,o),
\end{align*}

where $\mathcal{C}_{Rn}(1937/1938,003)$ is the measurement in the reference years 1937/38 in object 003 and $\mathcal{C}_{Rn}(t_0(o_0(o)), o_0(o))$ denotes the radon exposure in the reference object $o_0$ of object $o$ in the reference year $t_0$ for this reference object. The assumed measurement errors are given as follows.

\begin{align*}
    C_{Rn}(t_0(o_0(o)),o_0(o)) & =\mathcal{C}_{Rn}(t_0(o_0(o)),o_0(o)) + U_{\mathcal{C},c}(t_0(o_0(o)),o_0(o))
\end{align*}
\begin{align*}
    C_{Rn}(1937/1938, 003) & =\mathcal{C}_{Rn}(1937/1938, 003)  + U_{\mathcal{C},c}(1937/1938, 003)  \\[6pt]
     f(p_{o,j}) & = \varphi(p_{o,j}) \cdot U_{\varphi,c}(p_{o,j})\\
      \varphi'(t,o,p_j) & = \varphi(p_{o,j}) \cdot U_{\varphi',B}(t,o,p_j) \\[6pt]
     w(p_t) & = \omega(p_t) \cdot U_{\omega,c}(p_t)\\
     \omega'(t,o) & = \omega(p_t) \cdot U_{\omega',B}(t,o)  \\[6pt]
     g(p_{t,o}) & = \gamma(p_{t,o}) \cdot U_{\gamma,c}(p_{t,o})\\
     \gamma'(t,o) & = \gamma(p_{t,o}) \cdot U_{\gamma',B}(t,o) \\[6pt]
     b(o) & = \mathcal{b}(o) \cdot U_{\mathcal{b},c}(o)\\
      \mathcal{b}'(t,o) & = \mathcal{b}(o) \cdot U_{\mathcal{b}',B}(t,o) \\[6pt]
     t_e(o) & = \tau_e(o) \cdot U_{\tau_e,c}(o)\\
      \tau_e'(t,o) & = \tau_e(o) \cdot U_{\tau_e',B}(t,o)
\end{align*}

\subsubsection*{M2\_Expert}
The calculation of the exposure is similar to model M2, except that no radon gas measurements are used, but expert estimations for the radon gas concentrations:

\begin{equation*}
    X_i(t, o, j)=\mathcal{C}_{Exp}(p_{t,o}) \cdot 12 \cdot \varphi^{\prime}(t, o, j) \cdot \omega^{\prime}(t, o) \cdot \gamma^{\prime}(t, o) \cdot l_i(t,o,j) \cdot \tau(t,o).
\end{equation*}

The measurement errors deviation only with respect to $\mathcal{C}_{Exp}(p_{to})$.

\begin{align*}
& C_{Exp}(p_{t, o}) =\mathcal{C}_{Exp}(p_{t, o}) \cdot U_{\mathcal{C_{Exp}}, c}(p_{t, o}) \\[6pt]
& f(p_{o, j}) =\varphi(p_{o, j}) \cdot U_{\varphi, c}(p_{o, j}) \\
& \varphi^{\prime}(t, o, j) =\varphi(p_{o, j}) \cdot U_{\varphi^{\prime}, B}(t, o, j) \\[6pt]
& w\left(p_t\right) =\omega\left(p_t\right) \cdot U_{\omega, c}\left(p_t\right) \\
& \omega^{\prime}(t, o) =\omega\left(p_t\right) \cdot U_{\omega{ }^{\prime}, B}(t, o) \\[6pt]
& g\left(p_{t, o}\right) =\gamma\left(p_{t, o}\right) \cdot U_{\gamma, c}\left(p_{t, o}\right) \\
& \gamma^{\prime}(t, o) =\gamma\left(p_{t, o}\right) \cdot U_{\gamma^{\prime}, B}(t, o) 
\end{align*}

\subsubsection*{M3}
M3 is also quite similar to model M2. Instead of radon gas measurements M3 directly uses radon progeny measurements, eliminating the need for an equilibrium factor. To account for disruptions in the ventilation systems, the ventilation correction factor $\varsigma'(t,o)$ is used for M3:

\begin{align*}
X_i(t,o,j) & = \mathcal{C}'_{RDP}(t,o) \cdot 12 \cdot   \varsigma'(t,o) \cdot \omega'(t,o) \cdot \varphi'(t,o,p_j) \cdot l_i(t,o,j) ,
\end{align*}

where the measurement error is assumed as follows.

\begin{align*}
C_{RDP}(p_{t,o}) & =\mathcal{C}_{RDP}(p_{t,o}) + U_{\mathcal{C},c}(p_{t,o}) \\
\mathcal{C}'_{RDP}(t,o) & =\mathcal{C}_{RDP}(p_{t,o}) \cdot U_{\mathcal{C}',B}(t,o) \cdot \tau(t,o) \\[6pt]
     f(p_{o,j}) & = \varphi(p_{o,j}) \cdot U_{\varphi,c}(p_{o,j})\\
     \varphi'(t,o,p_j) & = \varphi(p_{o,j}) \cdot U_{\varphi',B}(t,o,p_j) \\[6pt]
     w(p_t) & = \omega(p_t) \cdot U_{\omega,c}(p_t)\\
     \omega'(t,o) & = \omega(p_t) \cdot U_{\omega',B}(t,o) \\[6pt]
     c(o) & = \varsigma(o) \cdot U_{\varsigma,c}(o)\\
     \varsigma'(t,o) & = \varsigma(o) \cdot U_{\varsigma',B}(t,o)
\end{align*}

\subsubsection*{M4}

Model M4 is simpler as it consists of only two uncertain factors:

\begin{align*}
X_i(t,o,j) & =\mathcal{E}'(t,o) \cdot \varphi'(t,o,p_j) \cdot l_i(t,o,j) ,
\end{align*}

with the error structure as follows.

\begin{align*}
   E(p_{t,o}) & =  \mathcal{E}(p_{t,o}) \cdot U_{E, c}(p_{t,o})\\
   \mathcal{E}'(t,o) & =  \mathcal{E}(p_{t,o}) \cdot U_{E, B}(t,o) \cdot \tau(t,o)\\[6pt]
     f(p_{o,j}) & = \varphi(p_{o,j}) \cdot U_{\varphi,c}(p_{o,j})\\
     \varphi'(t,o,p_j) & = \varphi(p_{o,j}) \cdot U_{\varphi',B}(t,o,p_j)
\end{align*}

\subsubsection*{Miners with no measurement error}
Furthermore, workers may not be exposed to radon progeny. This was assumed for a miner $i$ at time $t$ if at least one of the following conditions is applicable:
\begin{enumerate}
    \item A miner worked in a surface object (even if the extended data of the Wismut cohort contained no value for the exposure, thus, even if WLM was missing in the data and not set to 0 WLM).
    \item The object and object type were unknown but radon exposure was specified as 0 WLM.
    \item The miner's activity weighting factor had a value of 0, which, when multiplied to a hewer's radon progeny exposure, also gives an exposure of 0 WLM.
\end{enumerate}

\newpage
\FloatBarrier

\section{Update scheme MCMC}
\label{sec:appendix:update_scheme_long}
Here, we describe the implemented update scheme in more detail. First, we show how the updates for a parameter of the disease model is conducted. Afterwards, we explain the update process for one of the uncertain factors as part of the latent exposure, that requires more customization.

\subsection{Updates for parameters of the disease model} \label{sec:updateLatentexposure}
For the sake of brevity, we only show the update for $\beta$. Updates for $\lambda$ can be done in a similar way. The update is like a step in the standard MH-algorithm. As we are looking at the parameters of the disease model, we can condition on the (cumulated) exposure and the parameters of the baseline hazard. Hence, the relevant part of the (unnormalized) posterior reduces to

\begin{equation*}
    \left[\beta \mid \boldsymbol\lambda, X^{\text {cum }}\right] \propto[\beta] \prod_{i, t}\left[Y_i \mid \boldsymbol\lambda, \beta, X_i^{\text {cum }}(t)\right].
\end{equation*}
At each iteration, a new state $\beta^\ast$ is proposed using a Gaussian with the old state as mean (or a truncated Gaussian for each $\lambda_k$ separately with support on $\mathbb{R}^+$). The new state is accepted with a probability according to the MH ratio. This is a standard procedure to sample from an arbitrary posterior distribution. See for example Brooks et al. \cite{brooks2011}.

\subsection{Update of the latent exposure} \label{sec:update_latent_expo}
Quantities that directly affect the latent exposure of a worker $X_i^{\text {cum}}(t)$ pose some challenges to the standard update of the algorithm: 1) The exposure is a time-varying variable that accumulates over time while it is constructed for each time point and worker individually. This fact has to be reflected within the survival model. 2) The dimension of the latent quantities does not align over Berkson and classical error components. To see this problem, consider the error structure of the working time factor again: While an independent classical error is assumed over all periods $p_t$ (and therefore shared over all other possibilities), the Berkson deviation is assumed to range over all time points $t$ \textit{and} objects $o$ as well. \\
These two points suggest a more complex update scheme. We show the used solution with respect to measurement model M2 and start with the cumulation of the exposure because it is more intuitive.\\

\subsubsection*{Cumulation of time-varying exposure}
Consider at one sampling step in the MCMC algorithm  a current state of all latent quantities. In principle, equation (\ref{eq:M2Berkson}) can be used to obtain individual exposures for worker $i$ at time $t$. The evaluation of the disease model within a MH-update requires cumulated exposure, such that a time-varying survival likelihood can be evaluated. We give a small example that may illustrate the characteristics of the $\Xicum$ for worker $i$ using only three workers where the first one received exposure for five years, the second one for two years and the last one for three years:
\begin{equation*}
    X^{cum}=\left(\begin{array}{l}
X_1(1) \\
X_1(1)+X_1(2) \\
X_1(1)+X_1(2)+X_1(3) \\
X_1(1)+X_1(2)+X_1(3)+X_1(4) \\
X_1(1)+X_1(2)+X_1(3)+X_1(4)+X_1(5) \\
X_2(1) \\
X_2(1)+X_2(2) \\
X_3(1) \\
X_3(1)+X_3(2) \\
X_3(1)+X_3(2)+X_3(3)
\end{array}\right),
\end{equation*}
where $X_i(t)$ is the individual exposure for a worker $i$ at time $t$. To achieve the cumulation in a fast way, we define a sparse cumulation matrix that holds a 1 at the relevant positions to cumulate values. The cumulated exposure from beforecan therefore be written as a matrix multiplication with a vector representing the individual exposures:
\begin{equation*}
  X^{cum} = \left(\begin{array}{cccccccccc}
1 &   &   &   &   &   &   &   &   &   \\
1 & 1 &   &   &   &   &   &   &   &   \\
1 & 1 & 1 &   &   &   &   &   &   &   \\
1 & 1 & 1 & 1 &   &   &   &   &   &   \\
1 & 1 & 1 & 1 & 1 &   &   &   &   &   \\
  &   &   &   &   & 1 &   &   &   &   \\
  &   &   &   &   & 1 & 1 &   &   &   \\
  &   &   &   &   &   &   & 1 &   &   \\
  &   &   &   &   &   &   & 1 & 1 &   \\
  &   &   &   &   &   &   & 1 & 1 & 1
\end{array}\right)\left(\begin{array}{c}
X_1(1) \\
X_1(2) \\
X_1(3) \\
X_1(4) \\
X_1(5) \\
X_2(1) \\
X_2(2) \\
X_3(1) \\
X_3(2) \\
X_3(3)
\end{array}\right).
\end{equation*}

Using a sparse matrix permits a fast and memory efficient cumulation of values as only non-zero values are used in the calculation. The matrix can be stored as a persistent object that is memory efficient and does not lead to computational overhead during the sampling process. This is an important step because the cumulation operation has to be done at every update of each uncertain factor separately to evaluate the likelihood within the MH-update.

\subsubsection*{Updating an uncertain factor across different domains}
The measurement error structure described in section \ref{sec:application} shows that a latent factor is possibly a combination of a classical and a Berkson error. A closer look reveals that the shared error structure for a Berkson and classical error may differ implying that the dimension of an uncertain factor is different for the Berkson and classical error part.  We call this the domain over which an error affects an uncertain factor. As the classical error affects rather the level of the exposure, its dimension is by definition less or equal to the Berkson error dimension. This dimension discrepancy poses a challenges to the update scheme. We show the update procedure for the working time factor exemplarily. The other factors are updated in a similar fashion with appropriate adaptions to its individual characteristics.
The error structure for the working time factor is again stated: 

\begin{equation*}
    w\left(p_t\right) =\omega\left(p_t\right) \cdot U_{\omega, c}\left(p_t\right) \quad \text{and} \quad  \omega^{\prime}(t, o) =\omega\left(p_t\right) \cdot U_{\omega{ }^{\prime}, B}(t, o).
\end{equation*}

In a first step, we propose new values for the level, i.e. $\omega\left(p_t\right)$ and calculate all required ratios for the classical error part. As the dimension raises when accounting for the Berkson error, we use a predefined mapping matrix to expand the dimension to fit the Berkson error structure, where values may repeat at the according domains (e.g. multiple years have the same value, that is defined by $p_t$). As follows, we explain it in more detail.

\subsubsection*{Classical error}
Given a current state of the latent mean value $\omega\left(p_t\right)$, a new state $\omega\left(p_t\right)^\ast$ is proposed from a log-normal distribution (note that the dimension of the state is defined over the domain of $p_t$). Using the new values, the measurement error is calculated via $U_{\omega, c}^\ast\left(p_t\right) = w(p_t)/\omega\left(p_t\right)^\ast$. Using the new values, one can calculate the parts of the MH-ratio that are concerned with the classical error: the classical measurement ratio, the exposure ratio and the proposal ratio. Everything is calculated on log-densities, to ensure faster evaluation and numerical stability.

\subsubsection*{Berkson error}
Given the new latent level values $\omega(p_t)^\ast$, we multiply them with a mapping matrix to obtain the required dimension for the Berkson error. Afterwards, new Berkson errors $U_{\omega^{\prime}, B}(t, o)^\ast$ are proposed and an element-wise multiplication is done to obtain $\omega^{\prime}(t, o)^\ast$. A simplified example illustrates the calculation: 

\begin{equation*}
    \left(\begin{array}{c}
\omega^{\prime}(1955,1) \\
\omega^{\prime}(1955,2) \\
\omega^{\prime}(1956,1) \\
\omega^{\prime}(1957,1) \\
\omega^{\prime}(1958,1) \\
\omega^{\prime}(1959,1) \\
\omega^{\prime}(1959,2) \\
\omega^{\prime}(1967,1) \\
\omega^{\prime}(1968,1) \\
\omega^{\prime}(1968,2)
\end{array}\right)^\ast 
=
\left(\begin{array}{lll}
1 &   &   \\
1 &   &   \\
1 &   &   \\
1 &   &   \\
1 &   &   \\
  & 1 &   \\
  & 1 &   \\
  &   & 1 \\
  &   & 1 \\
  &   & 1
\end{array}\right)\left(\begin{array}{l}
\omega(1) \\
\omega(2) \\
\omega(3)
\end{array}\right)^\ast \odot
\left(\begin{array}{c}
U_{\omega^{\prime}, B}(1955,1) \\
U_{\omega^{\prime}, B}(1955,2) \\
U_{\omega^{\prime}, B}(1956,1) \\
U_{\omega^{\prime}, B}(1957,1) \\
U_{\omega^{\prime}, B}(1958,1) \\
U_{\omega^{\prime}, B}(1959,1) \\
U_{\omega^{\prime}, B}(1959,2) \\
U_{\omega^{\prime}, B}(1967,1) \\
U_{\omega^{\prime}, B}(1968,1) \\
U_{\omega^{\prime}, B}(1968,2)
\end{array}\right)^\ast,
\end{equation*}

where we use '$\odot$' to denote an element-wise multiplication. In this  simplified example, the dimension for the classical error is three, as there exist three periods, i,e, $p_t \in \{1,2,3\}$. The Berkson error is shared across seven different years and two objects leading to a dimension of ten when considering the unique observed combination\footnote{Note, that it is not necessary, that all combinations of  years and objects must exist. For instance, in the shown example, $o=2$ is not relevant for t=1956}. Given the new proposed values, one can use them to calculate the (log-) proposal and measurement ratio for the Berkson part and add them to the MH-ratio.\\
Afterwards, the new values are mapped to the full dimension of latent exposure over all individuals using a second mapping matrix with appropriate dimension in the same manner as before. The new values are then used, to calculate a new state for the latent exposure:
\begin{equation} \label{eq:calcTrueXefficient}
    X_i^\ast(t, o, j)^{[\tilde i]} = X(t,o,j)^{[\tilde i]} \cdot \omega^{\prime}(t, o)^\ast / \omega^{\prime}(t, o).
\end{equation}
Here, we use squared brackets with an index $\tilde i$ in the superscript to denote, that the calculation is only done for rows where an exposure value is calculated using measurement model M2. Calculating the new exposure in that way minimizes the computational operations. Also note, that we do not use equation (\ref{eq:M2Berkson}) directly and just modify the old values.\\
Given the new potential state of the exposure, one can cumulate the values using a sparse cumulation matrix as described before and use it to calculate and add the part of the MH-ratio concerned with the disease model. 
Afterwards, the MH-ratio is calculated using the sum of the different sub-ratios and the new proposed state is accept or reject as in standard methodology. The full procedure is iterated for each of the latent factors of the model with adapted mapping matrices. Also note, that the calculation in (\ref{eq:calcTrueXefficient}) is solely used for measurement model M2 and is adapted for the other measurement models described in section \ref{sec:appendix:ME_models_all} of this supplementary material. In particular measurement model M1a requires a bit more work. Moreover, it is worth to note, that for workers, that were not exposed at certain time periods, we apply a measurement model M0 that is assumed to have no measurement error. In the implementation, we use a masked array to exclude them from the calculations. We refer to the openly available code for more details.\\

\newpage
\FloatBarrier

\section{Priors, distributional assumptions and measurement error magnitude}
\label{sec:appendix:priors}

This is an extensive list of all chosen prior distributions and the related hyper-parameters, including the magnitude of measurement error (standard deviations):

\subsubsection*{Prior distributions for the parameters of the disease model}

\begin{itemize}
    \item $\beta \sim N(0,100^2)$
    \item $\lambda_1 \sim Ga(600, 1/10000000)$
    \item $\lambda_2 \sim Ga(12000, 1/1000000)$ 
    \item $\lambda_3 \sim Ga(46000, 1/1000000)$
    \item $\lambda_4 \sim Ga(1000, 1/100000)$
\end{itemize}

\subsubsection*{Distributional assumptions for the exposure models}
\begin{itemize}
    \item $\mathcal{C}_{Rn}(t_0(o_0(o)),o_0(o)) \sim N^+(22.5, 4^2)$
    \item $\mathcal{C}_{Rn}(1937/1938, 003) \sim N^+(34.09, 10^2)$
    \item $\mathcal{b}(o) \sim B_{[0.15, 1.1]}(1, 1)$
    \item $\tau_e(o) \sim B_{[0.3, 1]}(1, 1)$ 
    \item $\mathcal{C}_{Rn}(p_{t,o}) \sim N^+(\mu_{\mathcal{C}_{Rn}}(t), \sigma^2_{\mathcal{C}_{Rn}}(t))$
    \item $\mathcal{C}_{Exp}(p_{t, o}) \sim N^+(\mu_{\mathcal{C}_{Exp}}(t), \sigma^2_{\mathcal{C}_{Exp}}(t))$
    \item $\mathcal{C}_{RDP}(p_{t,o}) \sim N^+(\mu_{\mathcal{C}_{RDP}}(t), \sigma^2_{\mathcal{C}_{RDP}}(t))$
    \item $\mathcal{\mathcal{E}}(p_{t,o}) \sim N^+(\mu_{\mathcal{E}}, \sigma^2_{\mathcal{E}})$
    \item $\zeta(o) \sim B_{[1, 1.7]}(a_\zeta, b_\zeta)$
    \item $\varphi(p_{o, j}) \sim B_{[0,1.3]}(a_\varphi, b_\varphi)$
    \item $\omega\left(p_t\right) \sim B_{[0.6,1.5]}(a_\omega, b_\omega)$
    \item $\gamma\left(p_{t, o}\right) \sim B_{[0.05,0.8]}(a_\gamma, b_\gamma)$
\end{itemize}

\subsubsection*{Prior distributions for the exposure models}
\begin{itemize}
    \item $\mu_{\mathcal{C}_{Rn}}(t) \sim N(6, 5^2)$ (one distribution for each year in M2)
    \item $\sigma_{\mathcal{C}_{Rn}}(t) \sim N(8,0.5^2)$ (one distribution for each year in M2)
    \item $\mu_{\mathcal{C}_{Exp}}(t) \sim N(1.78,3^2)$ (one distribution for each year in M2\_Expert)
    \item $ \sigma_{\mathcal{C}_{Exp}}(t) \sim N(0.79,2^2)$ (one distribution for each year in M2\_Expert)
    \item $\mu_{\mathcal{C}_{RDP}}(t) \sim N(0.15,0.03^2)$ (one distribution for each year in M3)
    \item $\sigma_{\mathcal{C}_{RDP}}(t) \sim N(0.2,0.03^2)$ (one distribution for each year in M3)
    \item $\mu_{\mathcal{E}_{Rn}}(t) \sim N(2,3^2)$ 
    \item $\sigma_{\mathcal{E}_{Rn}}(t) \sim N(0.8,2^2)$
    \item $a_\zeta \sim N(3,2^2)$
    \item $b_\zeta \sim N(3,2^2)$
    \item $a_\varphi \sim N(3,2^2)$
    \item $b_\varphi \sim N(3,2^2)$
    \item $a_\omega \sim N(3,2^2)$
    \item $b_\omega \sim N(3,2^2)$
    \item $a_\gamma \sim N(3,2^2)$
    \item $b_\gamma \sim N(3,2^2)$
\end{itemize}

\subsubsection*{Specification of measurement error magnitude}
Table \ref{tab:quantified_sds} provides an overview of the used standard deviations of the measurement error for the measurement error models. The table is an adapted version of the table in \cite{ellenbach2023ermittlung}. The table shows the for each parameter the involving measurement error model and the according error type: This is either a Berkson or a classical error type and either additive, implying a (truncated) normal distribution or a multiplicative error that is modeled using a log-normal distribution. The fourth column shows the quantified standard deviation, which is primarily based on a very limited set of information.

\begin{table}
\centering
\resizebox{0.85\columnwidth}{!}{%
\begin{tabular}{llll}
Uncertain factor                                                                    & \begin{tabular}[c]{@{}l@{}}Measurement \\model\end{tabular}                                                            & \begin{tabular}[c]{@{}l@{}} assumed \\ measurement \\ error type \end{tabular}                                                        & \begin{tabular}[c]{@{}l@{}}standard \\deviation\end{tabular} \\ 

\hline
\hline

$\mathcal{C}_{Rn}(t_0(o_0(o)), o_0(o))$                                                              & M1a                                                                                                               & \begin{tabular}[c]{@{}l@{}}$U_{\mathcal{C},c}$: classical \\additive\end{tabular}       & 5.29                                                         \\ 
\hline
$\mathcal{C}_{Rn}(1937/38,003)$                                                                   & M1a                                                                                                                    & \begin{tabular}[c]{@{}l@{}}$U_{\mathcal{C},c}$: classical \\additive\end{tabular}       & 6.56                                                         \\ 
\hline

\multirow{2}{*}{$\mathcal{b}(o)$}                                                           & \multirow{2}{*}{M1a}                                                                                                   & \begin{tabular}[c]{@{}l@{}}$U_{\mathcal{b},c}$: classical \\multiplicative\end{tabular} & 0.33                                                         \\
                                                                             &                                                                                                                        & \begin{tabular}[c]{@{}l@{}}$U_{\mathcal{b}',B}$: Berkson \\multiplicative\end{tabular}   & 0.69                                                         \\ 
\hline

\multirow{2}{*}{$\tau_e(o)$}                                                      & \multirow{2}{*}{M1a}                                                                                              & \begin{tabular}[c]{@{}l@{}}$U_{\tau_e,c}$: classical \\multiplicative\end{tabular} & 0.37                                                         \\
                                                                             &                                                                                                                        & \begin{tabular}[c]{@{}l@{}}$U_{\tau_e',B}$: Berkson \\multiplicative\end{tabular}   & 0.33                                                          \\
\hline

\multirow{2}{*}{$\mathcal{C}_{Rn}(p_{t,o})$}                                                      & \multirow{2}{*}{M2}                                                                                              & \begin{tabular}[c]{@{}l@{}}$U_{\mathcal{C},c}$: classical \\multiplicative\end{tabular} & 0.59                                                         \\
                                                                             &                                                                                                                        & \begin{tabular}[c]{@{}l@{}}$U_{\mathcal{C},B}$: Berkson \\multiplicative\end{tabular}   & 0.33                                                          \\
\hline

$\mathcal{C}_{Exp}(p_{t, o})$                                       & M2\_Expert                                                                                                    & \begin{tabular}[c]{@{}l@{}}$U_{\mathcal{C},c}$: classical \\additive\end{tabular}       & 0.936                                                         \\

\hline

\multirow{2}{*}{$\mathcal{C}_{RDP}(p_{t,o})$}                                                           & \multirow{2}{*}{M3}                                                                                                    & \begin{tabular}[c]{@{}l@{}}$U_{\mathcal{C},c}$: classical \\additive\end{tabular} & 0.03                                                         \\
                                                                             &                                                                                                                        & \begin{tabular}[c]{@{}l@{}}$U_{\mathcal{C},B}$: Berkson \\multiplicative\end{tabular}   & 0.13                                                       \\ 

\hline

\multirow{2}{*}{$\varsigma(o)$}                                                           & \multirow{2}{*}{M3}                                                                                                    & \begin{tabular}[c]{@{}l@{}}$U_{\varsigma,c}$: classical \\multiplicative\end{tabular} & 0.33                                                         \\
                                                                             &                                                                                                                        & \begin{tabular}[c]{@{}l@{}}$U_{\varsigma',B}$: Berkson \\multiplicative\end{tabular}   & 1.45                                                       \\ 

\hline

\multirow{2}{*}{$\mathcal{E}(p_{t,o})$}                                                           & \multirow{2}{*}{M4}                                                                                                    & \begin{tabular}[c]{@{}l@{}}$U_{\mathcal{E},c}$: classical \\multiplicative\end{tabular} & 0.936                                                         \\
                                                                             &                                                                                                                        & \begin{tabular}[c]{@{}l@{}}$U_{\mathcal{E},B}$: Berkson \\multiplicative\end{tabular}   & 0.18                                                       \\ 

\hline

\multirow{2}{*}{$\omega(p_t)$}                                                           & \multirow{2}{*}{\begin{tabular}[c]{@{}l@{}}M1a, M2, \\ M2\_Expert, \\ M3\end{tabular}}                          & \begin{tabular}[c]{@{}l@{}}$U_{\omega,c}$: classical \\multiplicative\end{tabular} & 0.04                                                         \\
                                                                             &                                                                                                                        & \begin{tabular}[c]{@{}l@{}}$U_{\omega',B}$: Berkson \\multiplicative\end{tabular}   & 0.12                                                         \\ 
\hline
\multirow{2}{*}{$\varphi(p_{o,j})$}                                                           & \multirow{2}{*}{\begin{tabular}[c]{@{}l@{}}M1a, M2, \\ M2\_Expert, \\ M3\end{tabular}}                                                                                                   & \begin{tabular}[c]{@{}l@{}}$U_{\varphi,c}$: classical \\multiplicative\end{tabular} & 0.33                                                         \\
                                                                             &                                                                                                                        & \begin{tabular}[c]{@{}l@{}}$U_{\varphi',B}$: Berkson \\multiplicative\end{tabular}   & 0.69                                                         \\
\hline
\multirow{2}{*}{$\gamma(p_{t,o})$}                                                           & \multirow{2}{*}{\begin{tabular}[c]{@{}l@{}}M1a, M2, \\ M2\_Expert\end{tabular}}                              & \begin{tabular}[c]{@{}l@{}}$U_{\gamma,c}$: classical \\multiplicative\end{tabular} & 0.23                                                         \\
                                                                             &                                                                                                                        & \begin{tabular}[c]{@{}l@{}}$U_{\gamma',B}$: Berkson \\multiplicative\end{tabular}   & 0.69                                                        
\end{tabular} %
}
\caption{Uncertain parameters with the measurement models in which they occur, their assumed error types and the quantified standard deviations}
\label{tab:quantified_sds}
\end{table}

\newpage
\FloatBarrier

\section{Convergence diagnostics application}
\label{sec:appendix:convergence_application}
\begin{figure}[ht]
    \centering
    \makebox[\textwidth][c]{\includegraphics[width=1.3\textwidth]{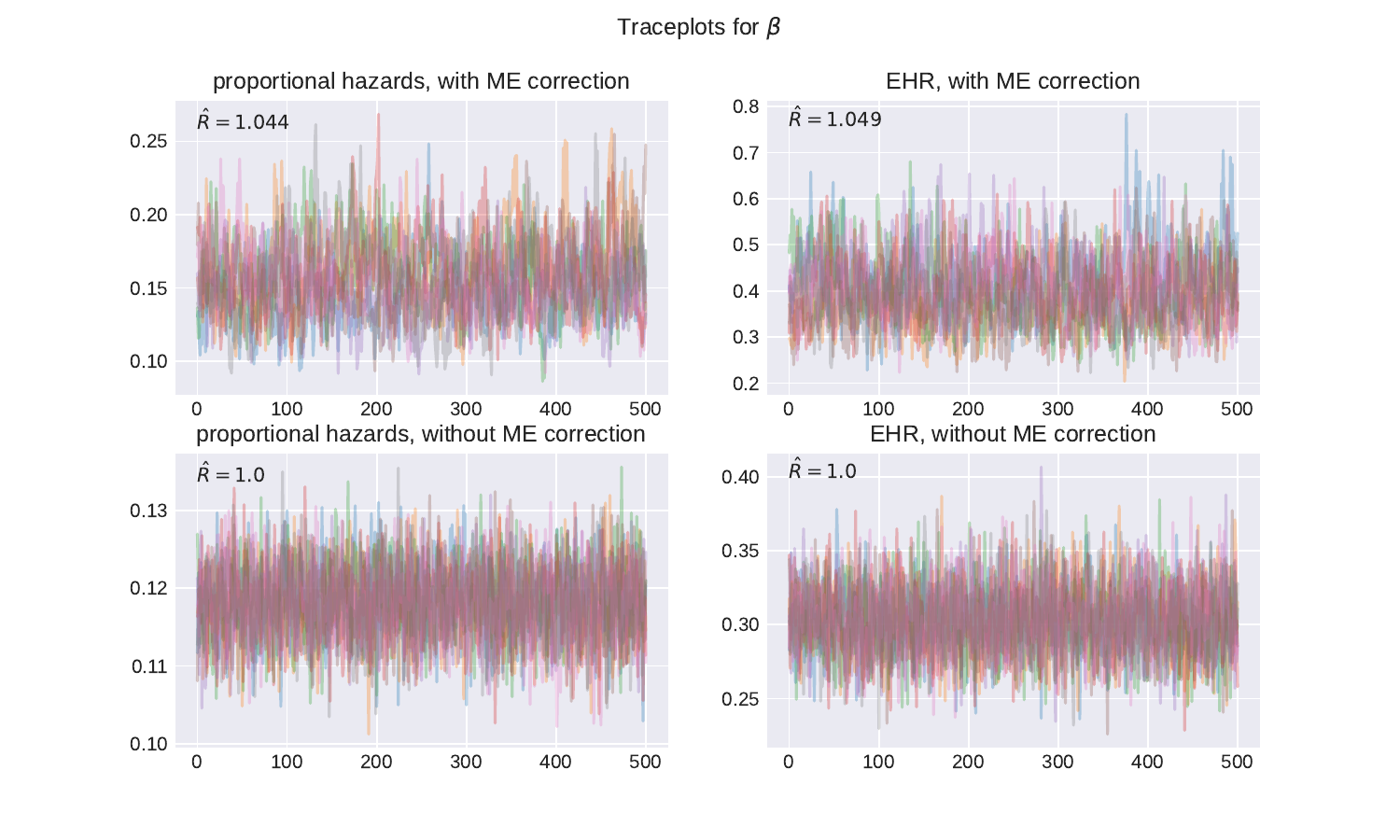}}
    \caption{Trace plots of eight generated Markov chains. The calculated $\hat{R}$ for each setting is shown in the top left corner of each plot.}
    \label{fig:convergence_application}
\end{figure}

Figure \ref{fig:convergence_application} shows the sampled eight Markov chains on the data of the Wismut cohort along with the calculated $\hat{R}$. The values indicate convergence. However, the values are relatively near the standard threshold of 1.05. The most simplest solution would be to run longer chains, that would lead to higher computational cost.

\newpage
\FloatBarrier

\section{Convergence diagnostics in the  simulation study}
\label{sec:appendix:results_simulation_study}

\begin{figure}[ht]
    \centering
    \makebox[\textwidth][c]{\includegraphics[width=1.05\textwidth]{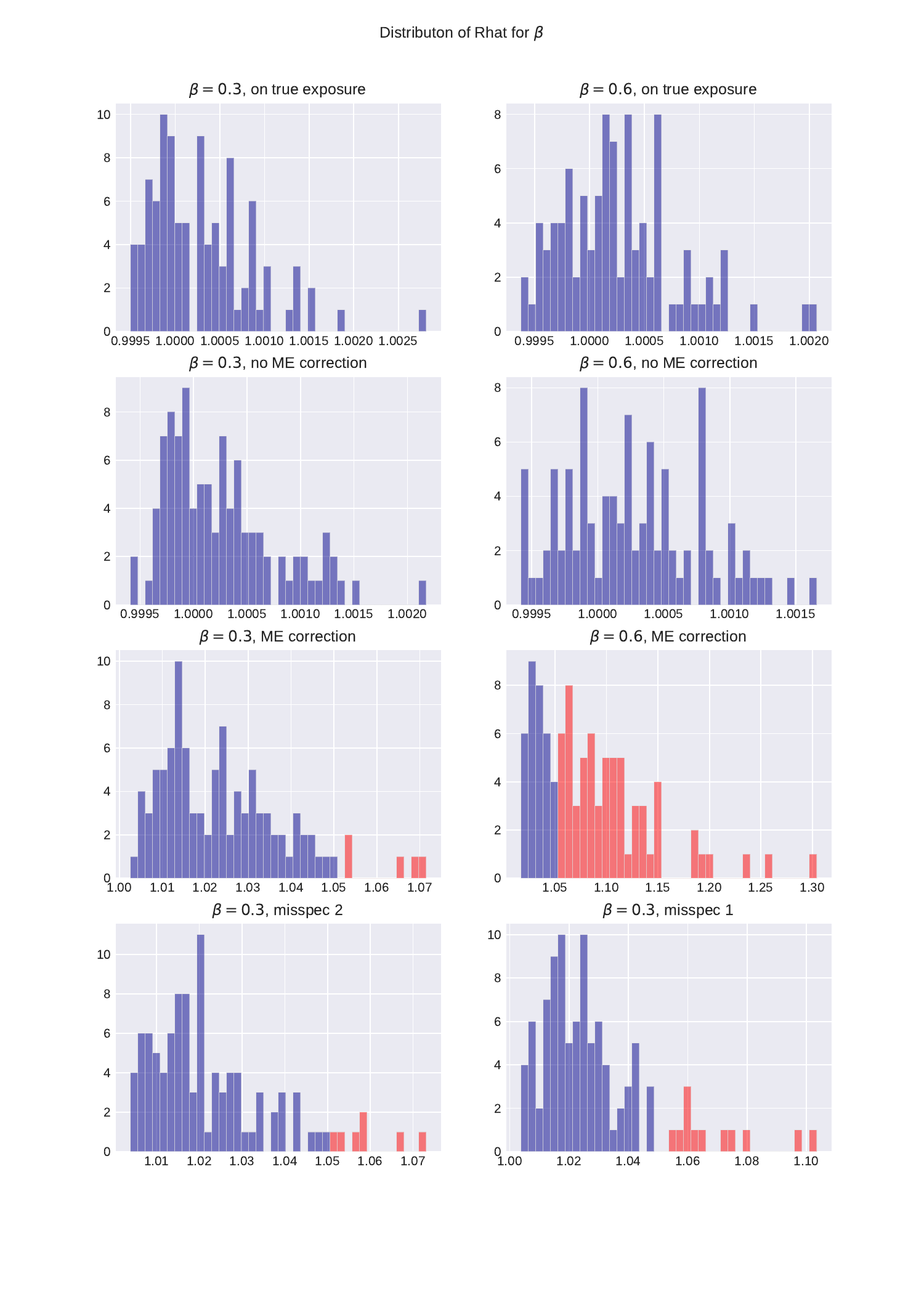}}
    \caption{Distribution of $\hat{R}$ over 100 simulated data sets. Each was calculated using four chains. Red indicates a  value $\ge 1.05$ implying non-convergence.}
    \label{fig:simulation_rhats}
\end{figure}

Figure \ref{fig:simulation_rhats} shows the distribution of $\hat{R}$ \citep{gelman1992inference, vehtari2021rank} for the different simulation scenarios and the true exposure over the 100 data sets, where we use four chains for the calculation of $\hat{R}$. It can be seen that for $\beta=0.6$, a significant share of the chains are indicated as non-converged (denoted in red, with a value of $\ge 1.05$). There are two potential reasons for that: 1) We used a limited number of iterations to keep computational cost feasible. The simulations were conducted with a lot less samples than the application. This was necessary because we repeated each run 100 times. 2) The quality of the simulated data for $\beta = 0.6$ is low. The simulation of survival times with time-varying exposure is a non-trivial task \citep{Montez2017}.  We experienced instable data generation: For such a high influence of the exposure on the survival time, the simulation gets even worse. This also aligns with the result on the true exposure: Even though the chains indicate convergence for $\beta = 0.6$, the distribution of the results was only calculated on 97 data sets. In three cases, the chains run into numerical issued and were therefore excluded afterwards. For comparison, we observe a mean value of the posterior of $\beta = 0.1553$ and even the 95\%-HDI level is with a value of $0.2032$ almost three times lower than the estimated value. For one of the 100 generated data sets it was not possible to sample numerical stable chains when accounting for measurement error.

\end{document}